\documentclass[a4paper,11pt]{article}
\pdfoutput=1 

\usepackage{jheppub} 

\usepackage[T1]{fontenc} 
\usepackage{graphicx}
\usepackage{amsmath}
\usepackage{amsmath}
\usepackage{subcaption}
\usepackage[utf8]{inputenc}

\preprint{\begin{flushright}
IP/BBSR/2018-4
 \end{flushright}}
\title{Long-lived Heavy Neutrinos from Higgs Decays}

\author[a]{Frank F. Deppisch,}
\author[a]{Wei Liu,}
\author[b,c]{Manimala Mitra}

\affiliation[a]{University College London, Gower Street, London WC1E 6BT, UK}
\affiliation[b]{Institute of Physics (IOP), Sachivalaya Marg, Bhubaneswar 751005, Odisha, India}
\affiliation[c]{Homi Bhabha National Institute, Training School Complex, Anushakti Nagar, Mumbai 400085,
India}

\emailAdd{f.deppisch@ucl.ac.uk}
\emailAdd{wei.liu.16@ucl.ac.uk}
\emailAdd{manimala@iopb.res.in}

\abstract{We investigate the pair-production of right-handed neutrinos via the Standard Model (SM) Higgs boson in a gauged $B-L$ model. The right-handed neutrinos with a mass of few tens of GeV generating viable light neutrino masses via the seesaw mechanism naturally exhibit displaced vertices and distinctive signatures at the LHC and proposed lepton colliders. The production rate of the right-handed neutrinos depends on the mixing between the SM Higgs and the exotic Higgs associated with the $B-L$ breaking, whereas their decay length depends on the active-sterile neutrino mixing. We focus on the displaced leptonic final states arising from such a process, and analyze the sensitivity reach of the LHC and proposed lepton colliders in probing the active-sterile neutrino mixing. We show that mixing to muons as small as $V_{\mu N} \approx 10^{-7}$ can be probed at the LHC with 100~fb$^{-1}$ and at proposed lepton colliders with 5000~fb$^{-1}$. The future high luminosity run at LHC and the proposed MATHUSLA detector may further improve this reach by an order of magnitude.}

\begin{document} 
\maketitle
\flushbottom
\section{Introduction}

The observation of light neutrino masses and mixing provides experimental evidence for the existence of physics beyond the Standard Model (SM). While the solar and atmospheric mass square differences and the mixing angles have been measured with considerably accuracy, the lightest neutrino mass and thus the neutrino mass scale, along with the nature of neutrinos, i.e. Dirac or Majorana, remains unknown. One of the simplest ultraviolet (UV) complete models to explain the light neutrino mass is the $U(1)_{B-L}$ model \cite{{Mohapatra:1980qe}}, where the vacuum expectation value (vev) of a SM singlet Higgs $\chi$ spontaneously breaks the $B-L$ symmetry and generates a Majorana mass for the heavy right-handed (RH) neutrinos $N_i$. Light neutrino Majorana masses are then generated via seesaw.

The model predicts the presence of RH neutrinos $N_i$, an additional gauge boson $Z^{\prime}$ that couples to the RH neutrinos as well as all other SM fermions, and an extra singlet-like Higgs state {$h_2$}. The $Z^{\prime}$ can be resonantly produced and it decays to a pair of RH neutrinos. Further decays of these heavy neutrinos produce lepton number violating signatures at the LHC \cite{Perez:2009mu}. The $Z^{\prime}$ also decays to SM fermions, leading to di-lepton and di-jet final states. While it can also decay to heavy neutrinos, with potentially spectacular displaced vertex signatures incorporating lepton flavour and number violating final states (see e.g. \cite{Deppisch:2013cya}), searches at the LHC for a heavy resonance decaying to lepton pairs puts stringent bounds on the $Z^{\prime}$ mass $M_{Z^{\prime}} > 4.5$~TeV \cite{Aaboud:2017buh} for a SM-like $Z^{\prime}$ state. {Similar searches on BOREXINO give an initial weak baseline limit for the $B-L$ breaking scale which is $M_{Z}^{\prime} / 2g^{\prime}$ $> $125 GeV, and LEP I yields $g' \lesssim 10^{-3}$ very close to the $Z$ resonance \cite{Batell:2016zod}, i.e. $M_{Z'} = m_Z$.} Further bounds from LEP-II constrain the $B-L$ breaking scale to be greater than 3.45~TeV \cite{Heeck:2014zfa, Cacciapaglia:2006pk, Anthony:2003ub, LEP:2003aa, Carena:2004xs}. The tight constraint on the $Z^{\prime}$ mass and the $B-L$ breaking scale considerably suppresses the production of RH neutrinos through the $Z^{\prime}$ mediated process. Other than this well addressed channel, the RH neutrinos can also be pair-produced from the BSM Higgs state {$h_2$}, as well as from the SM Higgs {$h_1$} for masses $2M_N < M_h$. For such low masses, the mixing of the RH neutrinos with the active neutrinos is $V_{lN} \lesssim 10^{-6}$ to produce the correct light neutrino masses. Such small couplings lead to sizeable decay lengths of heavy neutrinos and thus to potentially displaced vertices at colliders. If several heavy neutrino states exist, with mass splittings comparable or smaller than their widths, macroscopic oscillations may occur as well \cite{Antusch:2017ebe}.

A number of studies have been conducted in recent years on displaced vertex searches for heavy neutrinos for Type-I seesaw. At the LHC, several CMS searches \cite{CMS:2014hka, CMS:2015pca, CMS-EXO-12-037} have studied displaced vertices from long-lived neutral particles in a similar mass range with no events observed so far. On the other hand, the direct search for heavy neutrinos produced from a $W$ boson yields a constraint of order $V_{lN} < 10^{-2}$ \cite{Khachatryan:2015gha}. The recent 13 TeV search for tri-lepton on the other hand constrains the active-sterile mixing down to $V^2_{lN} \lesssim 10^{-5}$ \cite{Sirunyan:2018mtv}. The displaced lepton-jet final state from $W$ decay can be used to search for RH neutrinos giving a better constraint for $m_b \lesssim M_N \lesssim m_W$ \cite{Izaguirre:2015pga}. Long-lived sterile neutrinos can also be searched for at LHCb as described in \cite{Antusch:2017hhu}.

Invisible Higgs decays to exotic particles, including massive neutrinos, were first suggested in \cite{Shrock:1982kd}. The specific signal of pair-production of heavy long-lived neutrinos via a SM-like Higgs, $h_\text{SM} \to NN$, was considered in \cite{Maiezza:2015lza, Dev:2017dui} in the context of left-right symmetric models, in \cite{Caputo:2017pit} using an effective operator approach and in \cite{Accomando:2016rpc} within the $U(1)_{B-L}$ model we are using.\footnote{The mode $h_\text{SM}\to NN$ also occurs in the sterile neutrino case without an additional gauge coupling, but it is doubly penalized by the small active-sterile neutrino mixing; instead, the mode $h_\text{SM}\to \nu N$ is described in \cite{Das:2017zjc}.} Considering the extended Higgs sector associated with the $B-L$ breaking generating the heavy neutrino masses, other related processes are possible. For example, the decay $\chi\to NN$ of the extra Higgs in the given model can be analyzed. If both $\chi$ and $N$ are light enough, the decay chain $h\to\chi\chi\to 4N$ is also possible; an analogous mode was considered in \cite{Nemevsek:2016enw} within the minimal left-right symmetric model. On the other hand, if $\chi$ is light and its mixing with the SM Higgs is suppressed, it may also be long-lived \cite{Dev:2017dui}. At lepton colliders, the background for decay lengths between 10~$\mu$m and 249~cm in the detector is expected to be negligible \cite{Antusch:2016vyf} and the sensitivity on the active-sterile neutrino mixing is of the order $V_{lN} \approx 10^{-5}$ \cite{Antusch:2016ejd} in displaced vertex searches. For further discussions on displaced vertex signatures of RH neutrinos in different models, see \cite{Cottin:2018kmq, Helo:2018qej, Mandal:2017tab, Nemevsek:2018bbt, Lindner:2016bgg}. In the broader context, an extensive recent review of collider searches in seesaw models of neutrino mass generation can be found in~\cite{Cai:2017mow}. 

In this work, we explore the possibility to detect low mass RH neutrinos in the mass range $M_N \sim 10-60$~GeV through displaced vertex searches in the framework of gauged $U(1)_{B-L}$ model. Specifically, we aim to estimate the sensitivity on the active-sterile neutrino mixing in light of the comparatively weak limits on the SM Higgs mixing with a singlet scalar. For this, we consider the pair-production of RH neutrinos from SM Higgs decay and study the detectability of leptonic final states. This particular production mode is not limited by a small active-sterile mixing, and can be considerably larger compared to the production via a very massive $Z^{\prime}$. The neutrino decay length, however, crucially depends on the mixing and for very low mixing, the RH neutrino will be long-lived. We focus on the ongoing LHC run-II, the future high-luminosity run of the LHC (HL-LHC), with and without the proposed detector option MATHUSLA \cite{Chou:2016lxi}, and the proposed future $e^+e^-$ colliders ILC and CEPC to probe the active-sterile mixing.

The paper is organized as follows: In Section~\ref{blreview}, we briefly review the $U(1)_{B-L}$ model setup. Following this, in Section~\ref{expcons}, we study the experimental constraints on the model parameters and in Section~\ref{pairprod}, we discuss generalities of the pair production of RH neutrinos through a SM Higgs. In the subsequent two sections, we analyze in detail the detection possibility of a low mass RH neutrino at the LHC and at future leptonic colliders, respectively, where we present a detailed simulation. In Section \ref{Sen}, we present the sensitivity reach of these colliders and we conclude in Section~\ref{cnclu}. 

\section{$B-L$ Gauge Model}
\label{blreview}
In addition to the particle content of the SM, the $U(1)_{B-L}$ model consists of an Abelian gauge field $B^\prime_\mu$, a SM singlet scalar field $\chi$ and three RH neutrinos $N_i$. The gauge group is $SU(3)_c\times SU(2)_L \times U(1)_Y \times U(1)_{B-L}$, where the scalar and fermionic fields $\chi$ and $N_i$ have $B-L$ charges $B-L = +2$ and $-1$, respectively. The scalar sector of the Lagrangian consists of
\begin{align}\label{Ls}
	{\cal L} \supset (D^{\mu}H)^\dagger(D_{\mu}H) 
	                       + (D^{\mu}\chi)^\dagger (D_{\mu}\chi) 
	                       - {\cal V}(H,\chi),
\end{align} 
where $H$ is the SM Higgs doublet and $V(H,\chi)$ is the scalar potential given by 
\begin{align}
\label{VHX}
	{\cal V}(H,\chi) = m^2 H^\dagger H + \mu^2 |\chi|^2 + \lambda_1 (H^\dagger H)^2 
	          + \lambda_2 |\chi|^4 + \lambda_3 H^\dagger H |\chi|^2.
\end{align}
Here, $D_\mu$ is the covariant derivative, 
\begin{align}
\label{DM}
	D_\mu = \partial_{\mu} + ig_{s}\mathcal{T}_\alpha G_\mu^\alpha 
	      + igT_a W_\mu^a + ig_1 Y B_\mu + i g_1^\prime Y_{B-L} B^\prime_\mu,
\end{align} 
where $G^\alpha_\mu$, $W^a_\mu$, $B_\mu$ are the usual SM gauge fields with associated couplings $g_s$, $g$, $g_1$ and generators $\mathcal{T}_\alpha$, $T_a$, $Y$. $B^\prime_\mu$ is the gauge field associated with the additional $U(1)_{B-L}$ symmetry with gauge strength $g_1^\prime$ and the $B-L$ quantum number $Y_{B-L}$. Consequently, the gauge sector of the model now includes the the kinetic term 
\begin{align}
\label{LYM}
	{\cal L} \supset
	-\frac{1}{4} F^{\prime\mu\nu} F_{\mu\nu}^\prime,
\end{align} 
with the field strength tensor of the $B-L$ gauge group $F^\prime_{\mu\nu} = \partial_\mu B^\prime_\nu - \partial_\nu B^\prime_\mu$. Note that we do not consider a mixing between the Abelian hypercharge and $B-L$ gauge fields for the minimal $B-L$ model. 

The fermion part of the Lagrangian now contains a term for the right-handed neutrinos
\begin{align}
\label{Lf}
	{\cal L} \supset
    i\overline{\nu_{Ri}}\gamma_\mu D^\mu \nu_{Ri}, 
\end{align} 
but is otherwise identical to the SM apart from the covariant derivatives incorporating the $B-L$ gauge field and the charges $Y_{B-L} = 1/3$ and $-1$ for the quark and lepton fields, respectively. Here, a summation over the fermion generations $i = 1, 2, 3$ is implied. Finally, the Lagrangian contains the additional Yukawa terms
\begin{align}
\label{LY}
	{\cal L} \supset 
	   - y_{ij}^\nu \overline{L_i}\nu_{Rj}\tilde{H}
	   - y_{ij}^M \overline{\nu^c_{Ri}} \nu_{Rj}\chi 
	   + \text{h.c.},
\end{align} 
where $L$ is the SM lepton doublet, $\tilde{H} = i\sigma^2 H^\ast$ and a summation over the generation indices $i,j = 1, 2, 3$ is implied again. The Yukawa matrices $y^\nu$ and $y^M$ are a priori arbitrary; the RH neutrino mass is generated due to breaking of the $B-L$ symmetry, with the mass matrix given by $M_R = 2 y^M \langle\chi\rangle$. The light neutrinos mix with the RH neutrinos via the Dirac mass matrix $m_D = y^\nu v$. The complete mass matrix in the $(\nu_L^c, \nu_R)$ basis is then
\begin{align}
\label{MD}
	{\cal M} = 
	\begin{pmatrix}
		0   & m_D \\
		m^T_D & M_R
\end{pmatrix},
\end{align} 
where 
\begin{align}
\label{MDM}
	m_D = y^\nu v, \quad M_R = 2 y^M \tilde{x}. 
\end{align} 
Here, $v = \langle H^0 \rangle \approx 176$~GeV and $\tilde{x} = \langle\chi\rangle$ are the vacuum expectation values for electroweak and $B-L$ symmetry breaking, respectively. In the seesaw limit, $M_R \gg m_D$, the light and heavy neutrino masses are 
\begin{align}
\label{seesaw}
	m_\nu \sim - m_D M^{-1}_R m^T_D, \quad M_N \sim M_R.
\end{align}
The flavour and mass eigenstates of the light and heavy neutrinos are connected as  
\begin{align}
\label{Neutrino}
	\begin{pmatrix}
		\nu_L^c \\ \nu_R
	\end{pmatrix} = 
	\begin{pmatrix}
		V_{LL} & V_{LR} \\
		V_{RL} & V_{RR}
	\end{pmatrix}
	\begin{pmatrix}
		\nu^c \\ N
	\end{pmatrix},
\end{align} 
schematically writing the 6-dimensional vectors and matrix in terms of 3-dimensional blocks in generation space. The mixing and the light neutrino masses are constrained by oscillation experiments to yield their observed values, i.e. the SM charged current lepton mixing $V_{LL} = U_\text{PMNS}$ (apart from small non-unitarity corrections and assuming the charged lepton mass matrix to be diagonal). For the case of one generation of a light and heavy neutrino we will consider in turn, this reduces to the $2\times 2$ matrix form 
\begin{align}\label{Rotation}
	\begin{pmatrix}
		\nu_L^c\\ \nu_R
	\end{pmatrix} = 
	\begin{pmatrix}
    	\cos\theta_\nu & -\sin\theta_\nu \\
		\sin\theta_\nu &  \cos\theta_\nu
	\end{pmatrix}
	\begin{pmatrix}
		\nu^c \\ N
	\end{pmatrix}.
\end{align}
For simplicity, we thus neglect mixing among flavours and therefore generations decouple. The Yukawa coupling matrix then becomes diagonal and we can write ($i = e, \mu, \tau$)
\begin{align}
\label{YY}
	y^\nu_{ii} = \frac{M_{N_i} V_{iN}}{v},
\end{align} 
using the neutrino seesaw relation. Here, $V_{iN}$ represents the active-sterile mixing, $\sin\theta_i = V_{iN}$, in the three generations, $V_{e N}$, $V_{\mu N}$, $V_{\tau N}$. 

Similar to the light and heavy neutrinos, the additional scalar singlet $\chi$ also mixes with the SM Higgs. The mass matrix of the Higgs fields $(H, \chi)$ at tree level is \cite{Robens:2015gla}
\begin{align}
\label{mass}
	M_h^2 = \begin{pmatrix}
		2\lambda_1 v^2 & \lambda_3 \tilde{x} v \\
		\lambda_3 \tilde{x} v & 2\lambda_2 \tilde{x}^2
	\end{pmatrix}.
\end{align}
The physical masses of the two Higgs $h_1, h_2$ are then 
\begin{align}
\label{Higgsmass}
	M^2_{h_{1(2)}} = \lambda_1 v^2 + \lambda_2 \tilde{x}^2 
			     -(+) \sqrt{(\lambda_1 v^2 - \lambda_2 \tilde{x}^2)^2 + (\lambda_3 \tilde{x}v)^2},
\end{align} 
and the physical Higgs states ($h_1,h_2)$ are related to the gauge states ($H, \chi$) as
\begin{align}
\label{Higgs mixing}
	\begin{pmatrix}
		h_1 \\ h_2
	\end{pmatrix} = 
	\begin{pmatrix}
    	\cos\alpha & -\sin\alpha \\
		\sin\alpha &  \cos\alpha
	\end{pmatrix}
	\begin{pmatrix}
		H \\ \chi
	\end{pmatrix}.
\end{align} 
The directly measurable parameters are the masses $M_{h_1}$ and $M_{h_2}$, as well as the mixing angle $\alpha$,
\begin{align}
\label{lambda}
	\tan(2\alpha) = \frac{\lambda_3 v \tilde{x}}{\lambda_2 \tilde{x}^2 - \lambda_1 v^2}.
\end{align}
In our subsequent analysis, we consider the Higgs $h_1$ to be SM-like with mass $m_{h_1} = 125$~GeV, while the other state $h_2$ is heavier. Other than the heavy neutrinos and this additional Higgs state, the model also has an extra gauge boson with mass $M_{Z^\prime} = 2 \tilde{x} g_1^\prime$. In the next section, we discuss the experimental constraints, specifically on the active-sterile mixing $V_{iN}$, and on the Higgs mixing angle $\alpha$.

\section{Experimental Constraints}
\label{expcons}

Here, we review the experimental constraints on model parameters - in particular, the RH neutrino mass $M_N$, the active-sterile mixing $V_{lN}$ (neglecting generational dependence), the Higgs mixing angle $\sin\alpha$ and the $B-L$ breaking scale $\tilde{x}$.

In the present work, we consider relatively low mass RH neutrinos, 1~GeV$ < M_N < M_{h_1}/2 = 62.5$~GeV in order to pair-produce them from SM Higgs decays. The RH neutrino mass and its mixing with the active neutrinos are tightly constrained assuming successful neutrino mass measurements. In a pure Type-I scenario with $B-L$ gauge symmetry, the light neutrino mass $m_\nu \sim \frac{m^2_D}{M_R} \sim V^2_{lN}M_N$ with  $V_{lN} \sim m_D/M_R$. The sub-eV scale light neutrino mass constraints from $0\nu\beta\beta$ and Tritium beta decay experiments as well as from cosmological observations such as Planck \cite{Ade:2015xua} fixes the active-sterile mixing, 
\begin{align}
	V_{lN} \approx 10^{-6}\sqrt{\frac{m_\nu / (0.1~\text{eV})}{M_N / (50~\text{GeV})}}.
\end{align}
However, note that for models such as an inverse seesaw, the active-sterile mixing can be significantly larger, $V_{lN} \sim 0.01$, only limited by direct searches (see e.g. \cite{Deppisch:2015qwa} and references therein), while still satisfying the light neutrino mass constraint \cite{Dev:2012bd}.  

The SM singlet Higgs and its mixing angle $\alpha$ with the SM Higgs is constrained by perturbativity and  unitarity considerations \cite{Pruna:2011me}, setting an upper limit on $M_{h_2}$ as a function of the $B-L$ breaking scale $\tilde{x}$, $M_{h_2} \lesssim 2\sqrt{2\pi/3}\tilde{x}$. In our subsequent analysis, we consider a mass of $M_{h_2} = 450$~GeV, although it does not directly enter into our considerations. Additionally, direct searches at the LHC for a BSM Higgs signal further constrains the mixing $\sin\alpha \lesssim 0.35$ in the aforementioned mass range \cite{CMS:xwa}. An indirect constraint on the Higgs mixing angle $\sin^2 \alpha \lesssim 0.31 $ can also be obtained from  the measurement of SM Higgs decays into  a number of   SM final states \cite{Khachatryan:2014jba, Banerjee:2015hoa}. The  bound coming from SM Higgs signal strength measurement is valid  for all masses  of the BSM Higgs $h_2$. Precision measurements of the $W$ mass give a competitive bound on the mixing angle $\alpha \lesssim 0.3$ for a wide mass range $M_{h_2} \gtrsim 300$~GeV \cite{Lopez-Val:2014jva}. In the present work, we consider the value $\sin\alpha = 0.3$ to determine the maximal sensitivity on the neutrino mixing. 

Searches at LEP-II \cite{Cacciapaglia:2006pk, Anthony:2003ub, LEP:2003aa, Carena:2004xs} for a resonance constrain $Z'$ mass and gauge coupling, and thus the $B-L$ breaking scale $\tilde{x} \equiv \frac{M_{Z^\prime}}{2g_1^\prime} \geq 3.45$~TeV. Resonance searches in $pp\to Z^\prime \to l^+l^-$ bound the $Z^\prime$ mass to $M_{Z^\prime} \gtrsim 4.5$~TeV \cite{Aaboud:2017buh} with a SM-valued gauge coupling. Thus we choose $\tilde{x} = \frac{M_{Z^\prime}}{2g^\prime_1} = 3.75$~TeV, in agreement with the LEP and LHC bounds, where we consider $M_{Z^\prime} = 6$~TeV, $g_1^\prime = 0.8$ for definiteness.

In summary, we consider the following model parameters:
\begin{gather}
	M_N = 1 - 60\text{ GeV}, \quad V_{lN} = 10^{-9} - 10^{-2}, \nonumber\\
	M_{Z^\prime} = 6\text{ TeV}, \quad g_1^\prime = 0.8, \quad \tilde{x} = 3.75\text{ TeV}, \\
	M_{h_2} = 450\text{ GeV}, \quad \sin\alpha = 0.3. \nonumber\\
	\nonumber
\end{gather}

\section{Right-handed Neutrino Production and Decay}
\label{pairprod} 

\subsection{Pair-Production through Higgs Resonance}
We first consider the production of RH neutrinos through an $s$-channel SM-like Higgs, $pp \to h_1 \to NN$. We will assume that only one species of RH neutrinos is sufficiently light and that it dominantly couples to one lepton generation only. For more complicated scenarios with multiple RH neutrinos being sufficiently light, so that the SM-like Higgs can decay to different $N_i N_i$ pairs,  it will be possible to constrain the relevant mixing parameters $V_{eN_i}, V_{\tau N_i}$. The leading-order coupling of the SM-like Higgs $h_1$ with the two RH neutrino states is $\kappa_{h_1NN} = y^M \sin\alpha \cos^2 \theta_{\nu} \simeq y^M \sin\alpha =\frac{M_N}{\tilde{x}}\sin\alpha$, where $y^M$ is the associated Yukawa coupling that can be written in terms of the RH neutrino mass $M_N$ and $B-L$ breaking vev $\tilde{x}$. We have considered a small active-sterile neutrino mixing, that leads to $\cos \theta_\nu \sim 1$. In such a parametrization, the production cross-section is inversely proportional to $\tilde{x}^2$ and proportional to $M^2_N$, with the latter being bounded by the kinematic threshold $2M_N \le M_{h_1}$, and independent of the active-sterile mixing.

\begin{figure}[t!]
\centering
\includegraphics[width=0.48\textwidth]{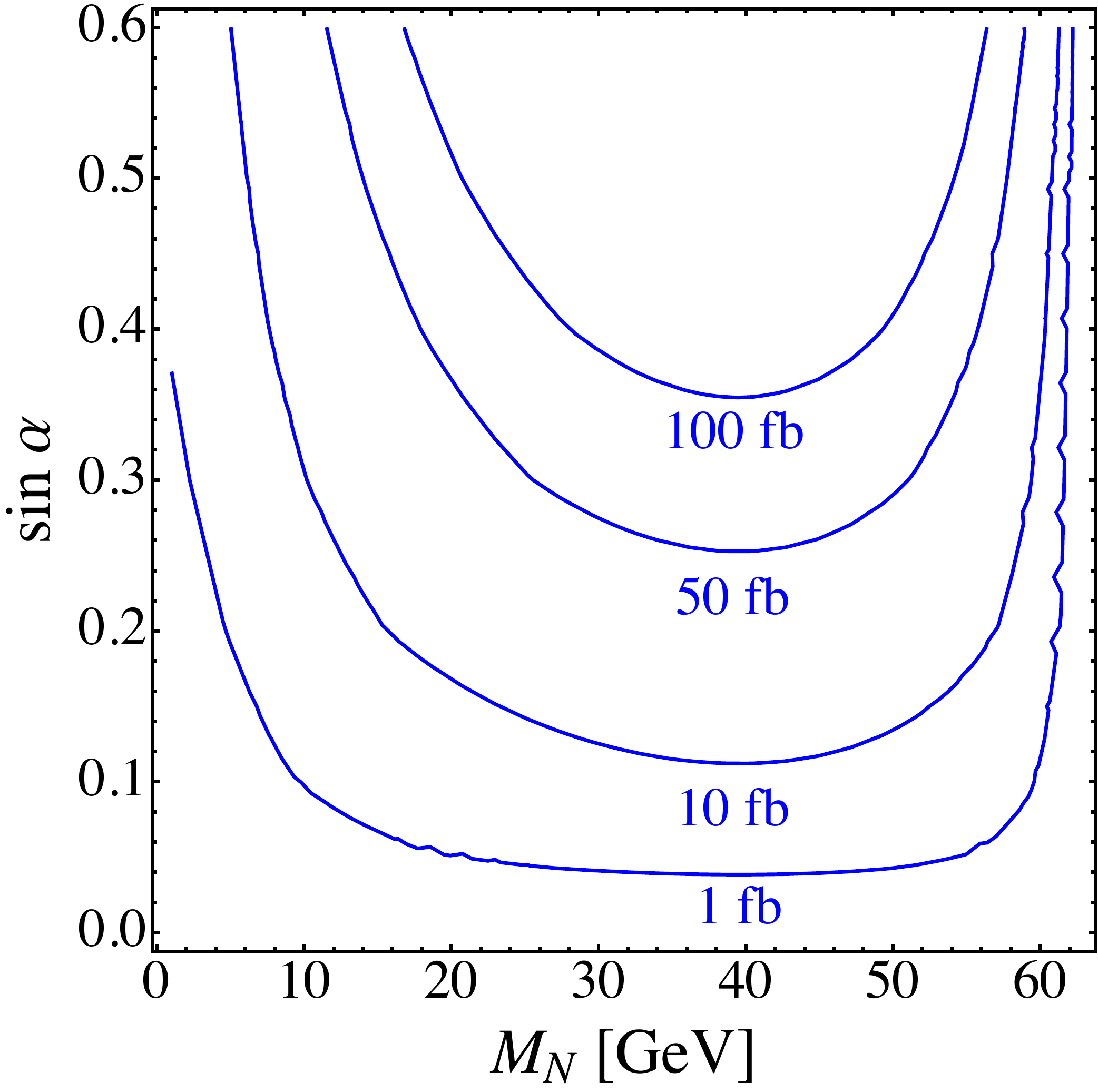}
\includegraphics[width=0.48\textwidth]{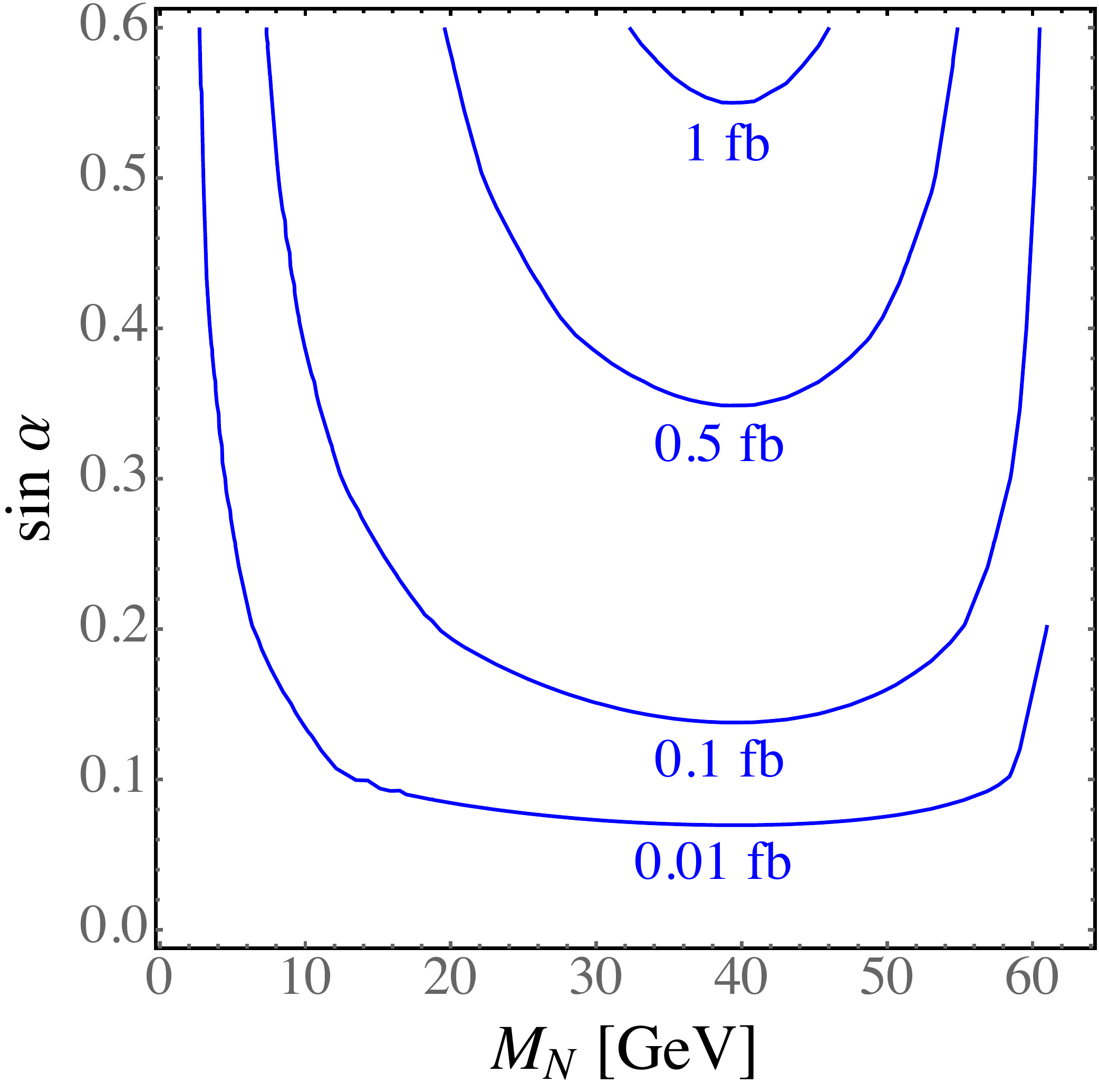}
\caption{Left: Cross section $\sigma(pp\to h_1 \to NN)$ as a function of the heavy neutrino mass $M_N$ and the Higgs mixing $\sin\alpha$ at the LHC with a center-of-mass energy of $\sqrt{s} = 13$~TeV. Right: Cross section $\sigma(e^+e^- \to Z \to Z h_1 \to Z + NN)$ as a function of the same parameters at an electron-positron collider with a center-of-mass energy of $\sqrt{s} = 250$~GeV.}
\label{fig:runzpcs}
\end{figure}
Note that the SM-like Higgs $h_1$ couples with the SM fermions and gauge bosons as $\cos\alpha$, while the BSM Higgs $h_2$ interacts with the same final states as $\sin\alpha$. We do not consider the decays $h_1 \to Z^\prime Z^\prime$ and $h_1 \to Z^\prime {Z^\prime}^*$, as we take the $Z^\prime$ to be much heavier than the SM gauge bosons. The branching ratio of SM-like Higgs can then be approximated as \cite{Accomando:2016rpc},
\begin{align}
\label{prod}
	\text{Br}(h_1 \to NN) = 
		\frac{\Gamma(h_1\to NN)}{{\Gamma(h)}_\text{SM}\cos\alpha^2 
		+ \Gamma(h_1 \to NN)},
\end{align}
where $\Gamma(h)_\text{SM}$ $\approx$ 4.2 $\times$ $10^{-3}$~GeV is the total decay width of the SM Higgs and
\begin{align}
	\Gamma(h_1\to NN) = \frac{2}{3}\sin^2\alpha\frac{M_N^2}{\tilde{x}^2}\frac{m_{h_1}}{8\pi}
	\left(1 - \frac{4M_N^2}{m_{h_1^2}}\right)^{3/2}
\end{align}
is the partial width of the SM-like Higgs to the new exotic channel. Similarly, the 13~TeV LHC cross section for the production of the SM-like $h_1$ is
\begin{align}
\label{prodcrossh1}
	\sigma(pp\to h_1) 
	= \sigma(pp\to h)_\text{SM}\cos^2\alpha
	\approx \cos^2\alpha \,(44\pm 4)~\text{pb}.
\end{align}

In Fig.~\ref{fig:runzpcs}~(left) we show the total $N$ production cross section as a function of the heavy neutrino mass and the Higgs mixing angle. As noted before, as long as the active-sterile neutrino mixing is sufficiently small, the production cross section is not sensitive to it. For a particular RH neutrino mass $M_N$, the cross section rises with increasing value of the mixing angle $\alpha$. On the other hand, for a higher value of $M_N$ within the kinematic threshold, a cross section as large as $50$~fb can be obtained for $\sin\alpha \sim 0.3$, in accordance with the LHC bound \cite{Lopez-Val:2014jva}, where the limit was derived for the Higgs singlet extension, however also applicable for the $B-L$ model. The cross section rapidly drops for small $M_N$. 

For the electron-positron collider case, we consider a center-of-mass energy $\sqrt{s} = 250$~GeV. The dominant Higgs production process is Higgs-Strahlung, $e^+e^- \to Z^* \to Z h_1$. In the SM, the cross section is $\sigma\sim 240$~fb for $\sqrt{s} = 250$~GeV, reduced by the Higgs mixing angle, $\propto \cos^2\alpha$, in our scenario. Hence, as shown in Fig.~\ref{fig:runzpcs}~{right}, the total $N$ production cross section is about 200 times smaller than at the LHC.

\subsection{Right-handed Neutrino Decay}
\begin{figure}[t!]
\centering
\includegraphics[width=0.8\textwidth]{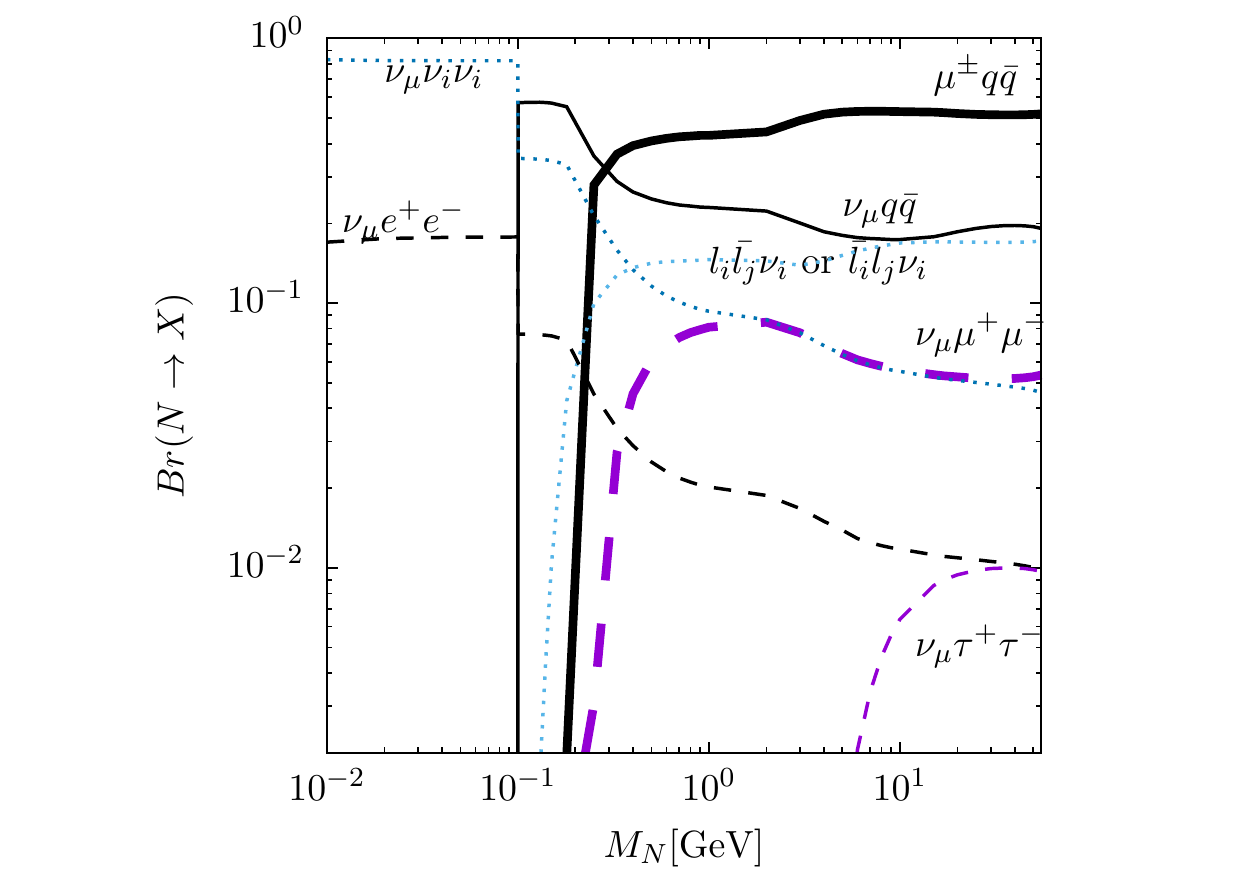}
\caption{Decay branching ratios $\text{Br}(N\to X)$ of the RH neutrino $N$ into the denoted channels $X$ as a function of $M_N$. Here, $i,j=e,\mu,\tau$ denotes lepton flavour with $i\neq j$ and the heavy neutrino is assumed to mix only with the light muon neutrino, $V_{\mu N}\neq 0$, $V_{e,\tau N} = 0$. In this case, the branching ratios are independent of $V_{\mu N}$.}
\label{fig:decaylength}
\end{figure}
The heavy neutrino, once produced, will decay to different SM states. For our mass region of interest, $M_N \lesssim 62.5$~GeV, and assuming there are no lighter exotic particles, the RH neutrino decays via three body processes such as $N\to \mu^\pm q\bar{q}$ and $N\to \mu^+\mu^-\nu_\mu$ for final states including muons. In this case and with our assumption that the heavy neutrino only mixes with one SM lepton generation, the branching ratios do not depend on the active-sterile mixing. In Fig.~\ref{fig:decaylength}, we show the branching ratios of the different decay modes as a function of the RH neutrino mass. In the region of interest, the heavy neutrino can decay to final states with one, two or three leptons. For relatively higher masses $M_N \gtrsim 1$~GeV, $N$ predominantly decays to $\mu q\bar{q}$, while for lower masses $M_N \lesssim 0.1$~GeV, the branching ratio of $N\to\nu\nu\nu$ becomes dominant. The given branching ratios, calculated using $MadGraph5aMC@NLO$~v2.5.5~\cite{Alwall:2014hca} and denoted by parton states correctly take into account decays to mesons for small RH neutrino masses. Approximately, the resulting decay length for $M_N \lesssim m_Z$ can be expressed as 
\begin{align}
\label{lengthapproxi}
	L_N \approx 0.025~\text{m} 
	\cdot \left(\frac{10^{-6}}{V_{\mu N}}\right)^2 
	\cdot \left(\frac{100~\text{GeV}}{M_N}\right)^5,
\end{align} 
where corrections due to the boost of parent $h_1$ will be discussed below. For a RH neutrino mass $M_N \approx 30$~GeV and a mixing $V_{\mu N} \approx 10^{-4}$, the decay length is of the order $L_N \approx 1$~mm. For smaller mixing, such as in the naive seesaw estimate $V_{\mu N} \approx 10^{-6}$, the decay length can be of the order of meters, potentially detectable through a displaced vertex at high energy colliders. For even smaller mixing and thus longer decay lengths, the decay products will escape the detector volume resulting in a missing energy signature.

\subsection{Estimate of Displaced Vertex Event Rate}
Before embarking on a detailed event simulation, we first estimate the rate of displaced vertex events from the production and decay of heavy neutrinos. As discussed in the previous subsection, we are interested in the decay length of the RH neutrino varying in the millimeter to several meters range, corresponding to active-sterile neutrino mixing in the relevant range for light neutrino mass generation. For long decay lengths, the heavy neutrino decay products will predominantly not register as prompt objects. Leptonic final states, such as a muon, or hadronic final states originating from such a RH neutrino decay can still be detected with displaced vertex searches at LHC \cite{ATLAS:2012av, Izaguirre:2015pga}. In the following, we estimate the number of such displaced events and thus the sensitivity of the 13~TeV LHC in probing the active-sterile mixing $V_{\mu N}$. 

We take into account the probability of the heavy neutrino decaying inside the detector and estimate the event rate corresponding to the observe displaced vertex events,
\begin{align}
\label{number}
	\frac{N_\text{events}}{\mathcal{L}} = 
	\sigma(pp\to h_1 \to NN)
	\times \text{Br}(N \to \text{final state}) 
	\times P(x_1 < x_N < x_2).
\end{align} 
Here, $P(x_1 < x_N < x_2)$ is the probability of the heavy neutrino decaying between distances $x_1$ and $x_2$,  taking into account the production mechanism, 
\begin{align}
\label{Pint}
	P(x_1 < x_N < x_2) = \int_0^\pi d\phi_N \int_0^1 d\beta_h \, 
	p(x_1 < x_N < x_2) f(\beta_h) g(\phi_N), 
\end{align} 
where $p(x_1 < x_N < x_2)$ represents the probability density of an individual neutrino to decay within the given range, $p(x_1 < x_N < x_2) = e^{-x_1/L^\prime_N} - e^{-x_2/L^\prime_N}$ (the primed decay length are in center-of-mass frame, not in the rest frame of the neutrino). $f(\beta_h)$ is the probability density function for the velocity of the SM-like Higgs $h_1$ at the LHC and $g(\phi_N)$ represents the probability density function for the production angle $\phi$ between the SM Higgs (i.e. the beam pipe) and the RH neutrino in the center-of-mass frame. 

\begin{figure}[t!]
\centering
\includegraphics[width=0.5\textwidth]{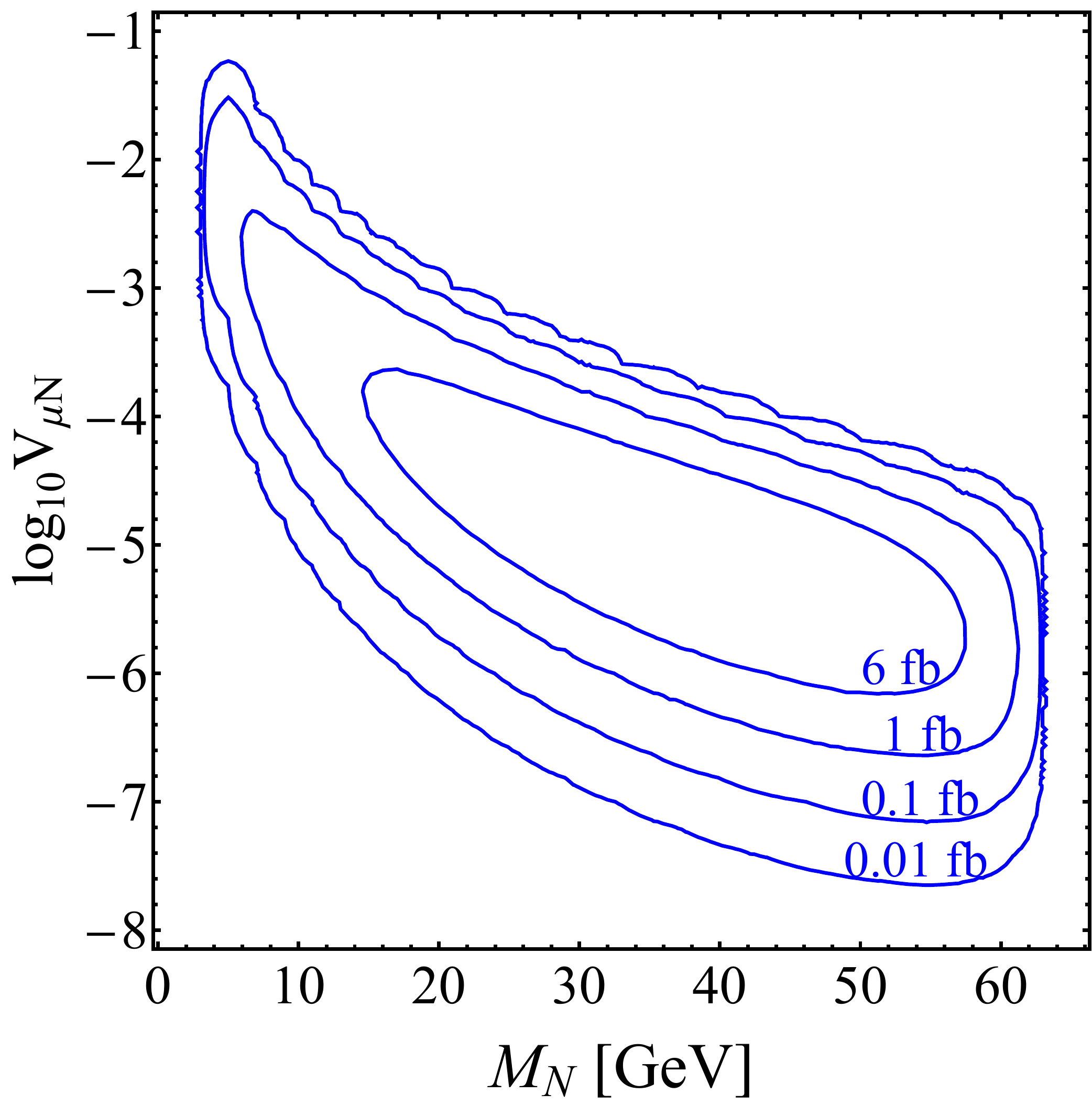}
\caption{Displaced vertex event rate at the LHC $\frac{N_\text{events}}{\mathcal{L}}(pp \to h_1 \to N N \to N \mu^\pm jj)$ with one $N$ decaying at a distance between 1~cm and 1~m, as a function of the neutrino mass and $M_N$ and the mixing $V_{\mu N}$. The Higgs mixing angle is set at $\sin\alpha = 0.3$.}   
\label{fig:length}
\end{figure}
We show the result of this analytic estimate of the rate of neutrinos that decay to a one muon final state within 1~cm and 1~m in Fig.~\ref{fig:length}, where we consider the displaced decay of one RH neutrino while we treat the second RH neutrino inclusively. In our subsequent analysis, we use Monte Carlo methods to fully simulate the events with a more detailed consideration of the detector geometry. Comparing the results in this section with the results in the next section, we find that both methods yield similar results.

\section{Displaced Vertex Event Simulation}
As discussed in the previous section, we focus on the RH neutrino decaying to leptonic final states. In particular, we focus on states with different $\mu$ multiplicity. This can be obtained assuming a diagonal mixing $V_{lN}$ or a mixing with hierarchies $V_{\mu N} \gg V_{eN}, V_{\tau N}$.  Below, we first present a brief discussion for the LHC and its future upgrades. A common, simplified detector layout and the details of the relevant detector parameters are shown in Fig.~\ref{fig:detector_diagrams} and Table.~\ref{tab:SiD}, respectively \cite{Accomando:2016rpc, Behnke:2013lya, Aihara:2009ad, CEPC-SPPCStudyGroup:2015csa}. The different variables we use for the geometry of detectors are as follows: $d_0$ is the transverse distance between the heavy neutrino $N$ and $\mu$, 
\begin{align}
\label{d0}
	|d_0| = |x p_y - y p_x|/p_T,
\end{align}
where $x$ and $y$ are the position where the right handed neutrino decayed, and $p_x$, $p_y$, $p_T$ are the components of momentum and transverse momentum of the final particles $\mu$, and $L_{xy}$ / $L_z$ are the transverse / longitudinal decay lengths of the RH neutrino, and $\sigma_d^t$ is the resolution of the detector in transverse distance.

Previous searches that analyzed the displaced decays of RH neutrinos are for example given in \cite{Izaguirre:2015pga, Accomando:2016rpc}. In \cite{Izaguirre:2015pga}, the authors propose to detect the RH neutrino having mass $M_N < M_W$ through a prompt lepton and a displaced lepton jet arising from the RH neutrino decay whereas in \cite{Accomando:2016rpc}, the authors propose to detect the RH neutrino through its displaced decays to at least a di-muon final state. Displaced decays of other exotic states have also been searched for at CMS and ATLAS in various physics scenarios, e.g. long-lived neutralinos \cite{Aad:2012zx} via one-muon and multi-jet final state; Higgs decaying to two long-lived particles \cite{CMS:2014hka, CMS:2015pca, CMS-EXO-12-037} producing two muons. Below, we describe briefly the proposed triggers and the estimated background, where we concentrate on the CMS and ATLAS detectors. We would like to mention though that long-lived particles may also be searched for at LHCb, see e.g. \cite{Aaij:2016xmb}

\subsection{LHC and Upgrade}
\begin{figure}[t!]
\centering
\includegraphics[width=0.5\textwidth]{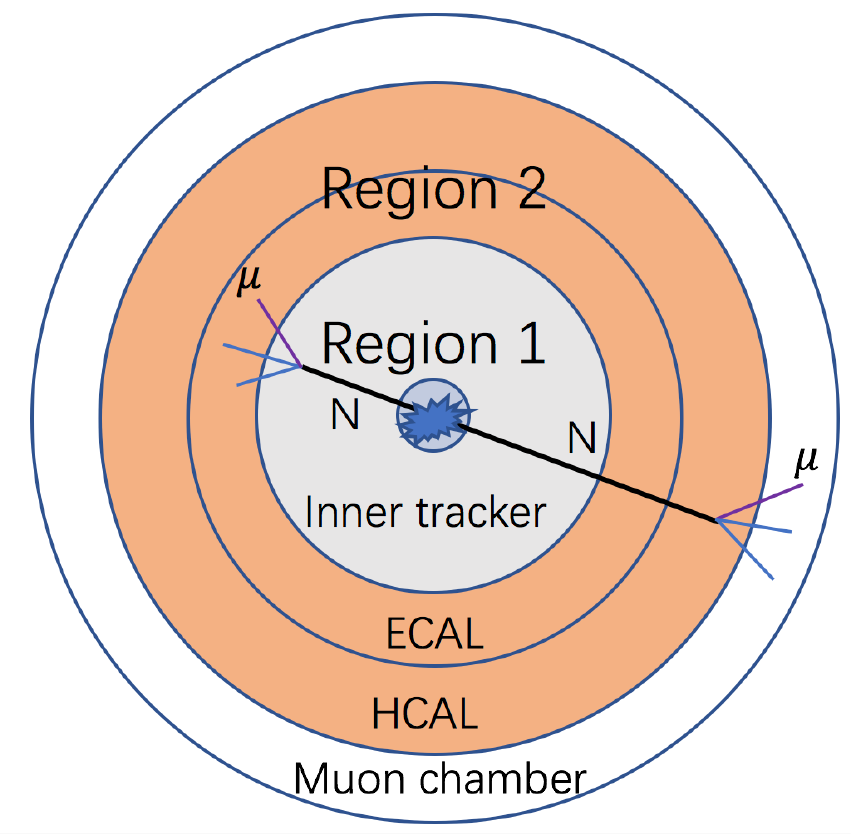}
\caption{Illustration of the simplified geometry of a typical detector we consider in our analysis. The innermost dark grey region is the vertex detector. The silicon tracker in light grey defines Region~1. The electromagnetic calorimeter (ECAL) and hadronic calorimeter (HCAL) outside the inner tracker and inside the muon chamber define Region~2.}  
\label{fig:detector_diagrams}
\end{figure}
We consider the RH neutrino decaying inside the tracker or muon chamber in the CMS detector \cite{Antusch:2016vyf}. We categorize the signal as $pp \to h_1 \to NN \to N\mu^\pm jj)$ (denoted as Channel~1) and $pp \to h_1 \to NN \to N\mu^+ \mu^-\nu_\mu)$ (denoted as Channel 2), respectively. We analyze the sensitivity reach in probing the active-sterile neutrino mixing $V_{\mu N}$ at the 13~TeV LHC using these channels. For the event simulation, we use $FeynRules$~2.3 \cite{Alloul:2013bka} with the model file {we created based on} \cite{Basso:2008iv}. The resulting Universal FeynRules Output (UFO) \cite{Degrande:2011ua} is fed to the Monte Carlo event generator $MadGraph5aMC@NLO$~v2.5.5~\cite{Alwall:2014hca}.

\paragraph{Triggers and Background} We differentiate between two regions, as shown in Fig.~\ref{fig:detector_diagrams} in the detector that can potentially detect displaced vertices \cite{Accomando:2016rpc}. Region~1 is chosen to probe long-lived RH neutrinos that decay within the inner tracker. The inner (grey) area of Fig.~\ref{fig:detector_diagrams} represents Region 1 which approximately consists of the inner tracker near the vertex detector such that RH neutrinos decaying within the inner tracker are registered via the tracks of final state muons. Region 2 is represented by the outer (orange) area which approximately consists of the electromagnetic (ECAL) and hadronic calorimeter (HCAL), and the inner region of the muon chamber. RH neutrino decays are registered in this region via tracks of the final state muons in the muon chamber \cite{Accomando:2016rpc}.

Our signal consists of a RH neutrino decaying with a displaced vertex to final states including muons, either $pp\to h_{1} \to NN \to N \mu^\pm jj$ or $N\mu^+\mu^-\nu_\mu$. While the branching ratios resulting in $N\mu^\pm jj$ are larger, the cuts needed to suppress the SM background are expected to be more stringent. In the literature \cite{Aad:2012zx, Izaguirre:2015pga, Accomando:2016rpc}, various selection criteria were employed for similar signatures. As described before, Ref.~\cite{Izaguirre:2015pga} uses the characteristic signal comprising of a prompt lepton (muon) and a heavy RH neutrino originating from a $W$ boson. Including the prompt muon and the particles from the RH neutrino decay, the signature consists of a muon jet, a reconstructed object with more than one muon track concentrated within a cone of radius $R_0$, $pp \to W^\pm \to \mu^\pm N \to \mu^\pm \mu^+\mu^-\nu_\mu$ or $\mu^\pm \mu^\pm j j$. The following cuts were used in this case: $p_T > 24 $~GeV for the prompt muon, the muon tracks inside the muon-jet should have $p_T > 6$~GeV, and the transverse impact parameter $d_0$ of the tracks in the muon-jet should satisfy 1~mm $< d_0 < 1.2$~mm. In our case,  the RH neutrino is pair-produced from a SM-like Higgs decay. Therefore, for   a  RH neutrino not too light,  decay products will generally  be un-collimated. Hence, the final state topology is different from that in \cite{Izaguirre:2015pga}. 

\begin{figure}[t!]
\centering
\includegraphics[width=0.5\textwidth]{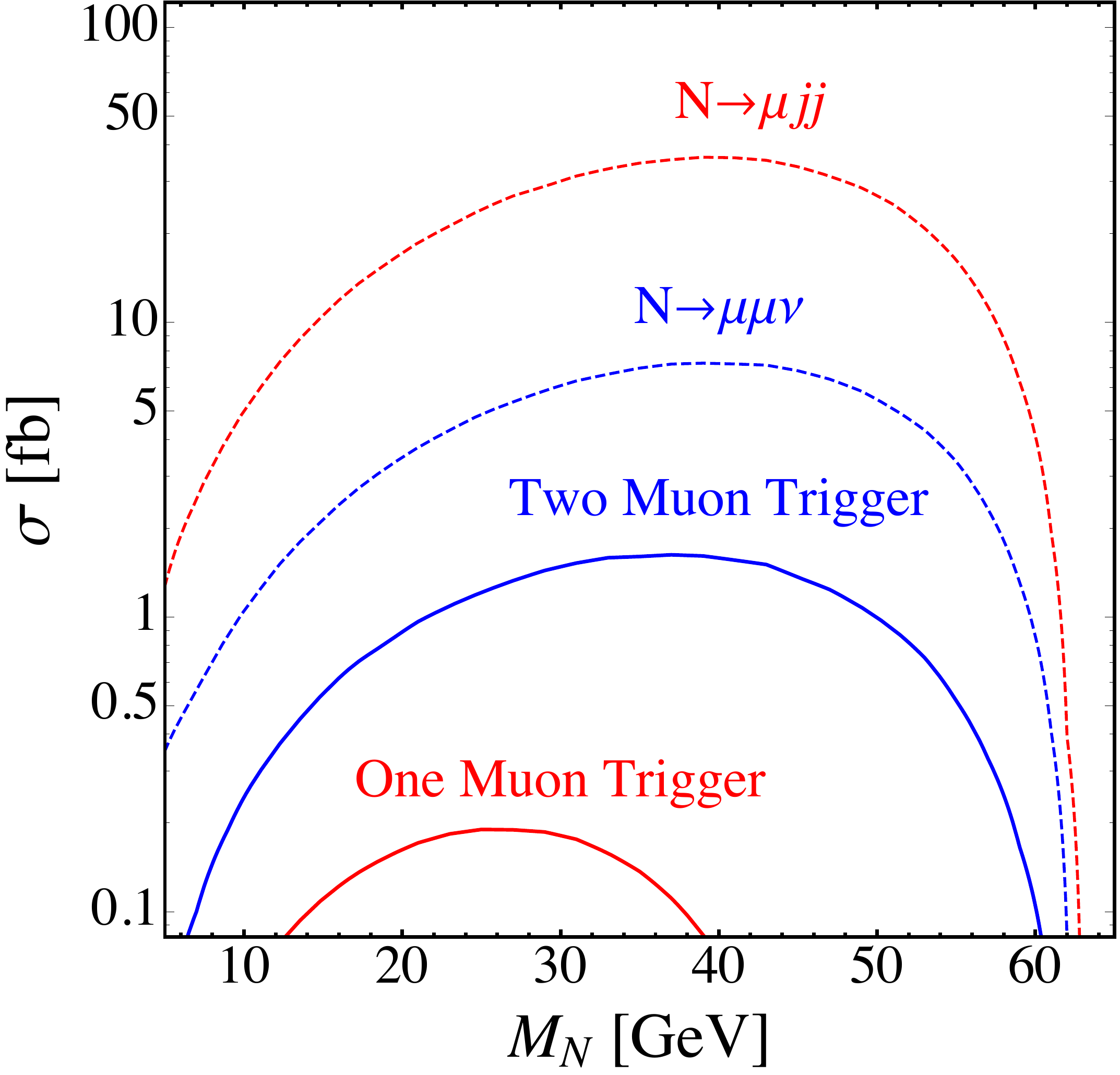}
\caption{Cross section for Channel~1 and Channel~2 as a function of the RH neutrino mass before (dashed) and after (solid) the corresponding kinematical cuts. The dashed red line represents the theoretical prediction for $\sigma(pp\to h_1 \to NN \to N \mu jj)$, while the solid red line corresponds to the cross section after the one muon selection, Eq.~\eqref{cutsonemuon}. Likewise, the dashed blue line gives $\sigma(pp\to h_1 \to NN \to N\mu\mu\nu )$ , and the solid blue line corresponds to the cross section after the two muon selection, Eq.~\eqref{cutsLHC}.}  
\label{fig:trigger}
\end{figure}

Ref.~\cite{Aad:2012zx} discusses a characteristic signal $\mu^\pm jj$, similar to our scenario, generated from displaced decays of a neutralino, which can be reinterpreted as our signal $pp\to h_1\to NN \to N\mu^\pm jj$. We refer to this as the one-muon trigger as there is a single muon in the final state. The selection criteria to identify a candidate both in the muon spectrometer and the inner detector are 
\begin{align}
\label{cutsonemuon} 
	p_T > 50\text{ GeV},\ 
	|\eta| < 1.07,\ 
	\triangle R = \sqrt{(\Delta\phi)^2 + (\Delta\eta)^2} < 0.15,\ 
	|d_0 | > 1.5\text{ mm}.
\end{align} 
Here, $\Delta\phi$ and $\Delta\eta$ are the differences between the azimuthal angle, and the  pseudo-rapidity of the muon identified by the trigger and that of the reconstructed muon, respectively. The cut on $\Delta R$ ensures, that the detected muon  corresponds to the muon identified by the trigger. Potential sources of background in this case are: i) random tracks that give rise to displaced vertices inside the beam-pipe, and ii) hadronic interactions with gas molecules that give rise to displaced vertices. A minimal invariant mass of 10~GeV of the tracks associated with the displaced vertex has been set. From the non-observation of displaced vertices, the number of corresponding background events was found to be $4^{+60}_{-4}$ $\times$ $10^{-3}$ for 4.4~fb$^{-1}$ at the 7~TeV LHC. As we make predictions for the 13~TeV LHC, background events should have higher $p_T$. With the above mentioned cuts, a significant amount of background will still be remaining.

In Ref.~\cite{Accomando:2016rpc}, relatively softer transverse momentum cuts have been used, with the same signal processes as in our case, $pp\to h_1 \to NN \to N\mu^+ \mu^- \nu_\mu$ and $N\mu^\pm jj$. The kinematic cuts used correspond to the presence of two muon tracks (Channel 2 with $\mu\mu\nu$ from one displaced vertex or Channel 1 with $\mu jj$ from two displaced vertices) that satisfy the following constraints on the transverse momentum of the leading ($\mu_1$) and sub-leading muon ($\mu_2$), pseudo-rapidity $\eta$ and isolation $\Delta R$ of the two tracks,
\begin{gather}
	p_T(\mu_1) > 26\text{ GeV}, ~~
	p_T(\mu_2) > 5\text{ GeV}, ~~
	|\eta| < 2.0 \nonumber\\
		|\Delta\Phi| < \pi/2,~~\Delta R > 0.2,~~
	\cos\theta_{\mu\mu} > -0.75.
	 \label{cutsLHC}
\end{gather}
An additional cut on the difference $\Delta\Phi$ in the azimuthal angle between the dilepton momentum vector and the vector from the primary vertex to the dilepton vertex has also been applied \cite{CMS:2014hka, CMS:2015pca, CMS-EXO-12-037}. Background due to cosmic ray muons can be rejected efficiently by correlating the corresponding hits with the beam collision time and with the cut on the angle between the muons, $\cos\theta_{\mu\mu}$ \cite{CMS:2015pca}.

Before attempting to reconstruct the displaced vertices geometrically, we first estimate the event rates using the above criteria Eqs.~\eqref{cutsonemuon} and \eqref{cutsLHC}. This is shown in Fig.~\ref{fig:trigger} demonstrating that although the RH neutrino has a higher branching ratio to the $\mu jj$ state, the high $p_T$ cut for Channel~1 ($\mu jj$) reduces the cross section by about a factor of ten whereas for Channel~2 ($\mu\mu\nu$) we use a relatively mild $p_T$ cut. 

In addition to the kinematical cuts, we also implement geometric cuts to reconstruct the displaced vertices. We use the following characteristics for Region~1 and 2 to represent a typical LHC detector \cite{Accomando:2016rpc} (also compare Table~\ref{tab:SiD}), where the variables used are discussed above,
\begin{align}
\label{cut1}
	\text{Region 1:}&\quad
	10~\text{cm} < |L_{xy}| < 50~\text{cm},\ |L_z| < 1.4~\text{m},\
	d_0/\sigma_d^t > 12,\ \sigma_d^t = 20~\mu\text{m}, \\
\label{cut2}
	\text{Region 2:}&\quad
	0.5~\text{m} < |L_{xy}| < 5~\text{m},\ |L_z| < 8~\text{m},\ 
	d_0/\sigma_d^t > 4,\ \sigma_d^t = 2~\text{cm}.
\end{align}
\begin{figure}[t!]
\centering
\includegraphics[width=0.5\textwidth]{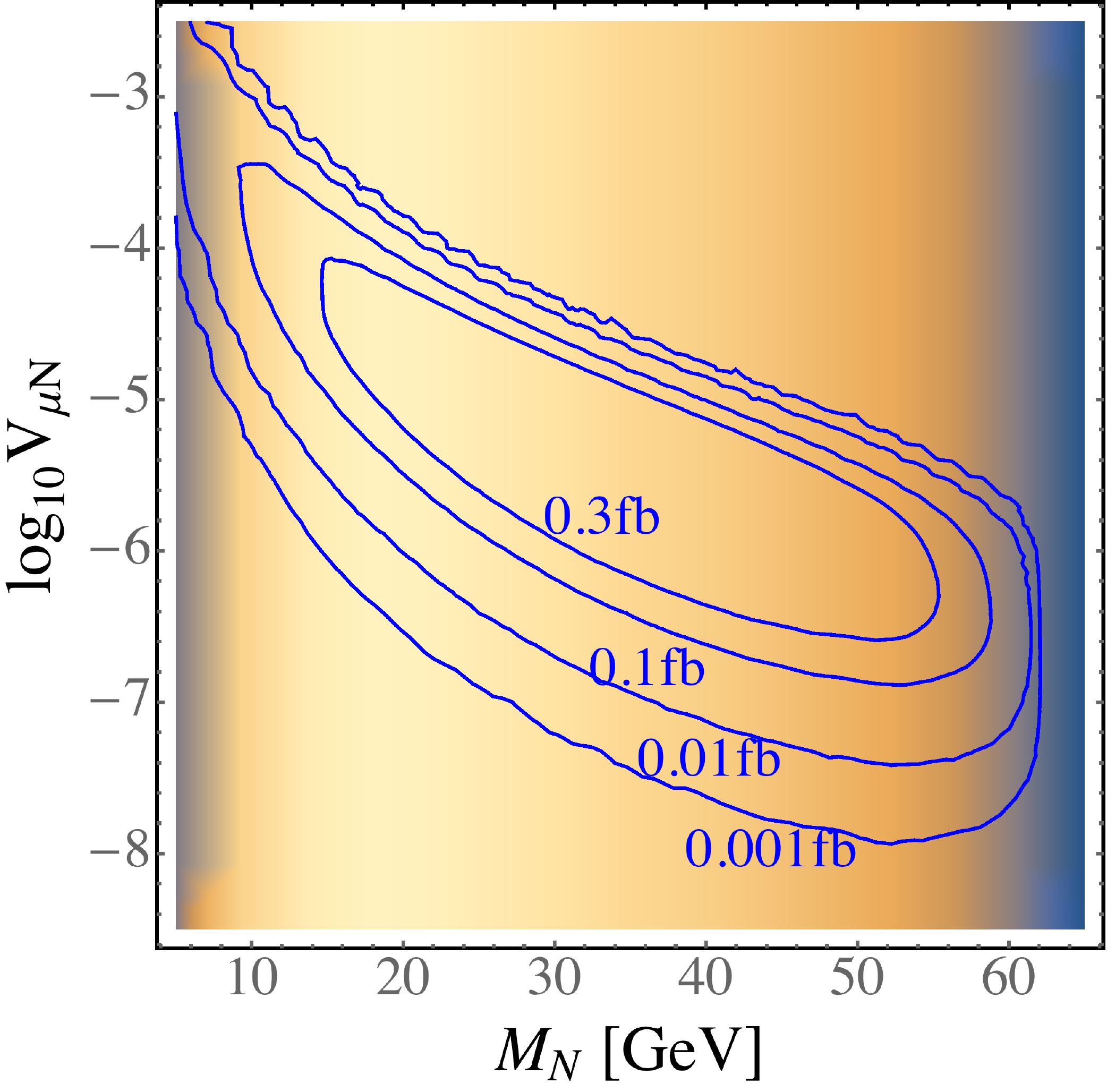}
\includegraphics[width=0.1\textwidth]{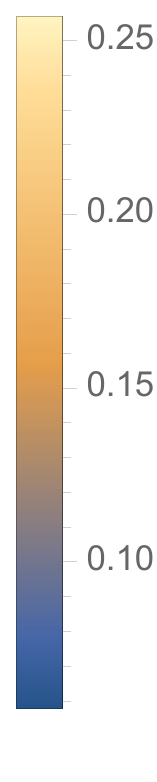}
\caption{Effective LHC displaced vertex cross section $\sigma(pp\to h_1 \to NN \to N\mu^+\mu^-\nu)$ as a function of the RH neutrino mass $M_N$ and the neutrino mixing. The background shading represents the kinematic efficiency $\epsilon_\text{kin}$ as indicated.}
\label{fig:LHC_newnew}
\end{figure}
The above kinematic and geometric selection cuts result in a reduction of the signal events as $\sigma\times \epsilon_\text{kin} \times \epsilon_\text{geo}$ with the kinematic and geometric efficiency $\epsilon_\text{kin}$ and $\epsilon_\text{geo}$, respectively. To demonstrate their impact, we show the resulting effective signal cross section for Channel~2 in Fig.~\ref{fig:LHC_newnew}. The kinematic efficiency is shown separately using the coloured shading. The maximal event rates after cuts can be as high as 0.3~fb which is not too dissimilar from Fig.~\ref{fig:length} considering the difference in the branching ratio of the final states.

\paragraph{Two Displaced Vertex Events}
In order to reduce the background further, we may demand to observe at least two muons from two different displaced vertices. When requiring two such displaced vertices, the signature is different and can contain 2, 3 or 4 muons altogether, i.e. $pp\to h_1 \to NN \to \mu^\pm jj, \mu^\pm jj$ which is the dominant channel, $pp \to h_1 \to NN \to \mu^\pm jj, \mu^-\mu^+\nu_\mu$ and $pp\to h_1\to NN \to \mu^-\mu^+\nu_\mu, \mu^-\mu^+ \nu_\mu$. With the requirement of two displaced vertices, the effective event rate due to the additional branching ratio and the geometric reconstruction efficiency is reduced by more than an order of magnitude. However this can be considered to be a cleaner selection to discover this specific model.

\subsection{MATHUSLA}
There are several proposals to equip the high luminosity run of the LHC (HL-LHC) with additional detectors to search for long-lived particles. As an example, we consider the proposal comprising of a large detector on the ground surface called MATHUSLA to detect ultra long-lived particles a few hundred meters away from the collision point \cite{Chou:2016lxi}. We estimate the sensitivity of this setup by applying one the following geometrical selection cuts:
\begin{gather}
	-100~\text{m} < L_x < 100~\text{m},\, 
	 100~\text{m} < L_y < 120~\text{m},\, 
	 100~\text{m} < L_z < 300~\text{m}, \nonumber\\
	 d_0/\sigma_d^t > 4,\,
	 \sigma_d^t = 2~\text{cm}.
\end{gather} 
Due to its setup, MATHUSLA has a comparatively small geometric coverage. However, it can potentially probe very small active-sterile neutrino mixing as it would be situated far away from the interaction point. 

\subsection{Future Electron-Positron Colliders}
\begin{table}[t!]
	\centering	
	\begin{tabular}{c|ccccc}
		\hline
		Region & Inner Radius & Outer Radius & z-Extent & $|d_0|/\sigma_d^t$ & $\sigma_d^t$ 
		\\ \hline
		LHC Region 1  & \phantom{0}10 & \phantom{0}50 & 140 & 12 & 0.02  \\
		LHC Region 2  & \phantom{0}50 & 500 & 800 &  4 & 2     \\\hline
		ILC Region 1  & \phantom{0}22 & 120 & 152 & 12 & 0.002 \\
		ILC Region 2  & 120 & 330 & 300 &  4 & 2     \\\hline
		CEPC Region 1 & \phantom{0}15 & 180 & 240 & 12 & 0.007 \\ 
		CEPC Region 2 & 180 & 440 & 400 &  4 & 2     \\\hline
	\end{tabular}
	\caption{Parameters of simplified detector geometries representing current and future detectors, namely LHC \cite{Chatrchyan:2008aa}, ILC \cite{Behnke:2013lya, Aihara:2009ad}, CEPC \cite{CEPC-SPPCStudyGroup:2015csa}. All length units are in cm.}
	\label{tab:SiD}
\end{table}
We are also interested in the sensitivity of proposed future electron-positron colliders. Leptonic colliders can benefit from a cleaner background, especially relevant for Higgs production. Here, we focus on the proposed International Linear Collider (ILC) and the Circular Electron Positron Collider (CEPC). We consider a center-of-mass energy of $\sqrt{s} = 250$~GeV and a luminosity of 5000~fb$^{-1}$. The dominant Higgs production process at an electron-positron collider is Higgs-Strahlung,  $e^+e^- \to Z^* \to Z h_1$. In the SM, the cross section is $\sigma\sim  240$~fb for $\sqrt{s} = 250$~GeV. Further suppression of the cross-section occurs  in our case, due to the Higgs mixing angle (see Fig.~\ref{fig:runzpcs}).

For the cuts on the kinematic variables we use 
\begin{eqnarray}
\label{cutsILC}
	p_T(l) > 10~\text{GeV},\ 
	|\eta| < 2.0,\ 
	\Delta R > 0.2,\ 
	\cos\theta_{\mu\mu} > -0.75.
\end{eqnarray} 
The selection criteria associated with displaced tracks are set analogous to the LHC analysis, as  described before. As for the detector type and geometry, the ILC proposes to use a Silicon Detector (SiD) \cite{Behnke:2013lya, Aihara:2009ad} for general purpose detection and precision measurements. In Table~\ref{tab:SiD}, we include the geometric parameters we use for the detectors of ILC and CEPC in our analysis \cite{Behnke:2013lya, Aihara:2009ad, CEPC-SPPCStudyGroup:2015csa}. For lepton colliders, we represent a silicon tracker as Region~1, and the components before the muon system as Region~2.

Due to the relatively smaller production cross section, the overall signal rate at the lepton colliders is smaller compared to LHC by about two orders of magnitude,  whereas the efficiency is larger. For the CEPC, due to the longer and larger detector compared to the ILC, the cross section of the CEPC is slightly larger. 

\section{Sensitivity Reach}
\label{Sen}
We now estimate the sensitivity of various colliders in probing the active-sterile neutrino mixing. We follow the approach discussed above and implement the kinematic and geometric cuts in a Monte Carlo simulation of our signal. We assume that the employed cuts and selection criteria remove the backgrounds at the LHC completely taking  first an optimistic view. This is justified, as the displaced vertex searches in  \cite{CMS:2014hka, CMS:2015pca, CMS-EXO-12-037} have observed no events. In addition, we will adopt a pessimistic view and take the upper limit on background events allowed by the non-observation and we scale it to the luminosity of 100~fb$^{-1}$ at the LHC and 3000~fb$^{-1}$ at the HL-LHC.

For the optimistic view assuming zero background events, following a Poisson distribution for the number of signal events, the lower and upper limits on the mean value of signal events $\mu$ are given by
\cite{Agashe:2014kda}:
\begin{align}
\label{poisson}
	\mu_\text{min} = \frac{1}{2} F^{-1}_{\chi^2}(  \alpha, 2n), \quad 
	\mu_\text{max} = \frac{1}{2} F^{-1}_{\chi^2}(1-\alpha, 2(n+1)),
\end{align} 
respectively. Here, $F_{\chi^2}(\alpha, n)$ is the cumulative distribution function for the $\chi^2$ distribution with $\alpha$ being the significance level and $n$ denoting the number of observed events. Probing a cross section with a sensitivity at 95\% C.L., the upper bound on $\mu = \sigma\times L$ for $n = 0$ is $3.09$ \cite{Agashe:2014kda}. Therefore, we consider a model parameter point with  $\mu > 3$ to be excluded at 95\% C.L. on non-observation of any event. 

For the pessimistic view, scaling the experimental upper limit on the background rate, we interpret the non-observation of displaced vertex events \cite{CMS:2014hka, CMS:2015pca, CMS-EXO-12-037} at 20.5~$fb^{-1}$, to yield an upper limit on the mean event rate  of 3 events. We scale this rate up for the 100~$fb^{-1}$ LHC and 3000~$fb^{-1}$ HL-LHC, giving 15 and 450 potential background events, respectively. However, we still consider leptonic colliders and MATHUSLA to be free from background. Therefore, we consider a parameter point to be excluded on non-observation if $\chi^2 = (N_\text{tot} - N_B)^2/ N_B > 3.84$ at 95~\% C.L. In the case of two separate displaced vertices, since this is a highly rare process, we think only the optimistic view is necessary.

\begin{figure}[t!]
\centering
\includegraphics[width=0.49\textwidth]{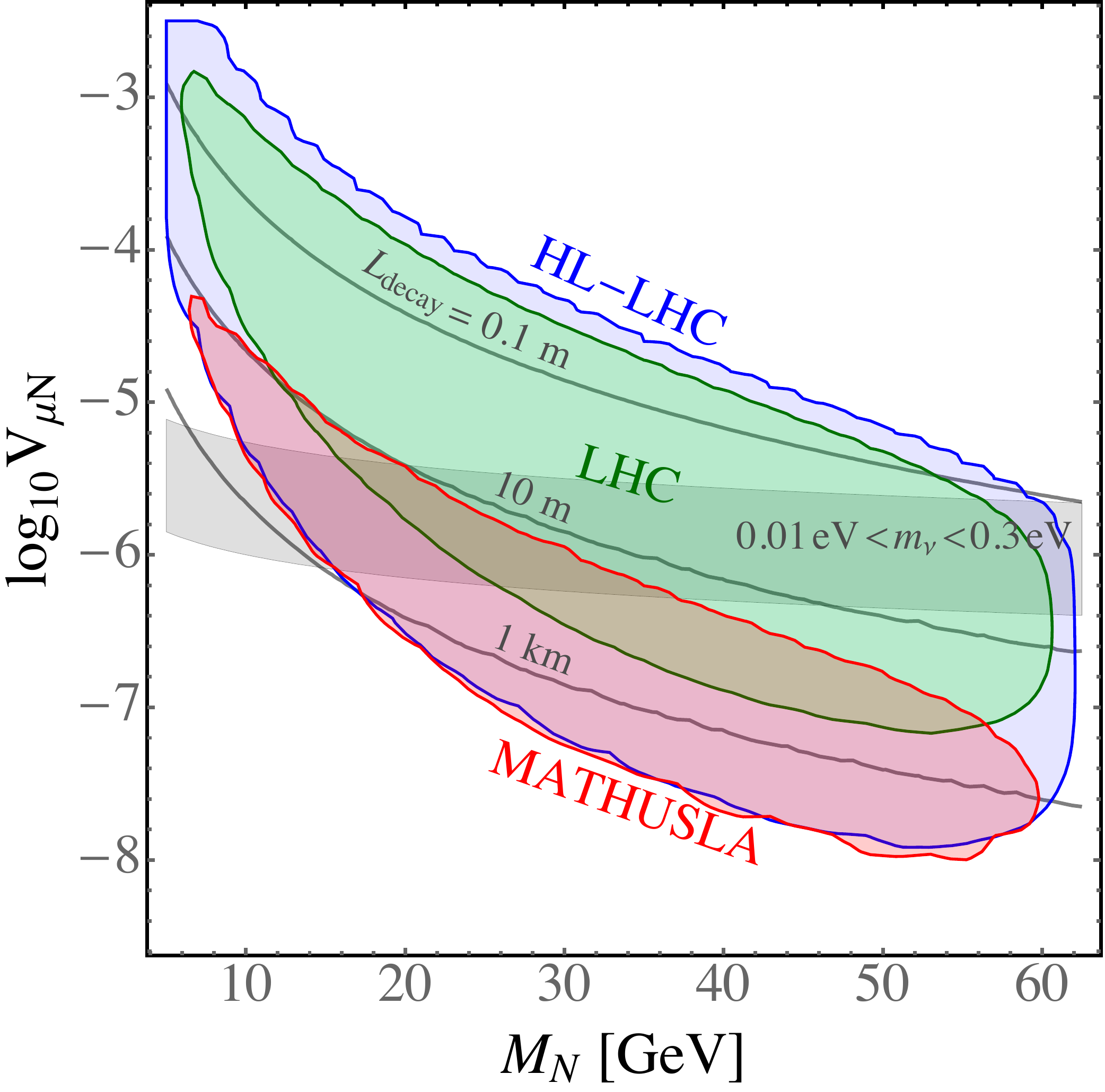}
\includegraphics[width=0.49\textwidth]{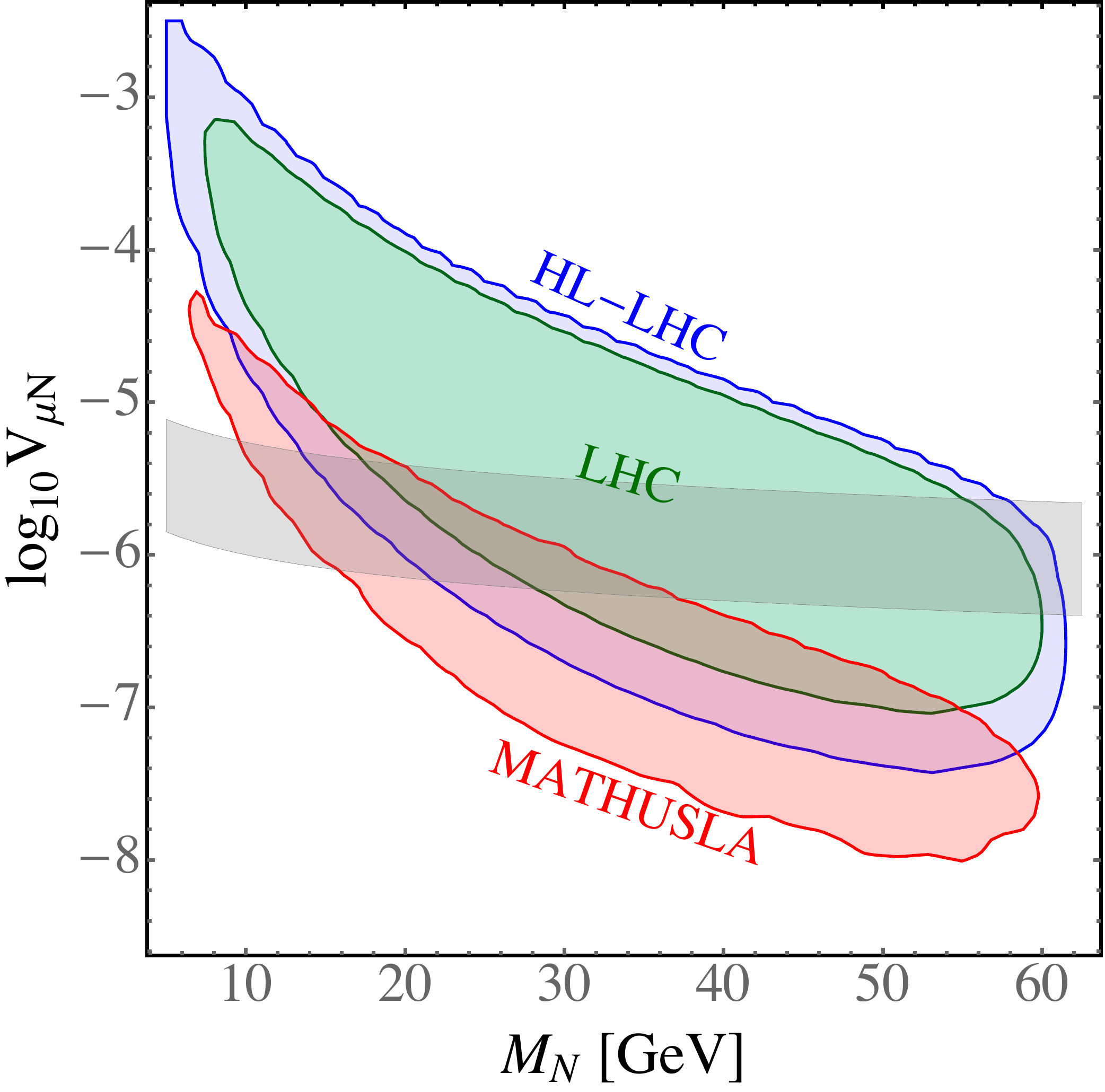}
\caption{Excluded regions in the $(M_N, V_{\mu N})$ parameter space at 95\%~C.L. assuming no observation of a single displaced vertex for the 100~fb$^{-1}$ LHC (green), the 3000~fb$^{-1}$ HL-LHC (blue) and the MATHUSLA option at HL-LHC (red). The left plot is in the  optimistic view assuming no background after selection criteria whereas the right plot is in the pessimistic view where the upper limit on the background rate from the non-observation at 20.5~fb$^{-1}$ is scaled to the different luminosities. The grey band indicates the parameter region where a light neutrino mass in the interesting range is generated, $0.01~\text{eV} < m_\nu = V_{\mu N}^2 m_N < 0.3$~eV. The red indicate proper decay lengths of the RH neutrino.}
\label{fig:LHC1DV}
\end{figure}
In Fig.~\ref{fig:LHC1DV} we show the resulting sensitivities at the LHC, HL-LHC (without and with MATHUSLA) in the optimistic view assuming no background (left) and the pessimistic view with a scaling of the background (right). We consider that one of the heavy neutrinos decays to muons in one displaced vertex. To estimate the sensitivity reach, we consider 100~fb$^{-1}$ luminosity for the LHC, 3000~fb$^{-1}$ for HL-LHC. Taking the optimistic and rather naive view of negligible background, it is evident that the 13~TeV LHC has a sensitivity reach down to $V_{\mu N} \approx 10^{-7}$ for RH neutrino masses around $M_N \approx 55$~GeV. The high-luminosity run of LHC can further probe smaller mixing angle, as low as $10^{-8}$ for a similar RH neutrino mass value. The sensitivity reach of MATHUSLA is similar to the HL-LHC; despite the much longer decay length being probed, the geometric coverage and hence effective cross section is smaller. We should stress again that in deriving these limit we considered a Higgs mixing $\sin\alpha = 0.3$, close to the experimental limit. However, future searches e.g. at leptonic collider will have a better sensitivity on the Higgs mixing $\sin\alpha \approx 0.06$ \cite{Gu:2017ckc}, resulting in a cross section about 30 times smaller. In the pessimistic view, Fig.~\ref{fig:LHC1DV}~(right), we see that due to the large background the HL-LHC can still reach a mixing as low as $10^{-7.5}$. Note that we assume that the background is constant at its upper limit for the whole parameter space; this is overly pessimistic as the background should get smaller as the RH neutrino becomes longer-lived because most SM background should have decayed away already. In both the left and right panel we assume the background for MATHUSLA to be negligible and the sensitivity regions are identical.

\begin{figure}[t!]
\centering
\includegraphics[width=0.49\textwidth]{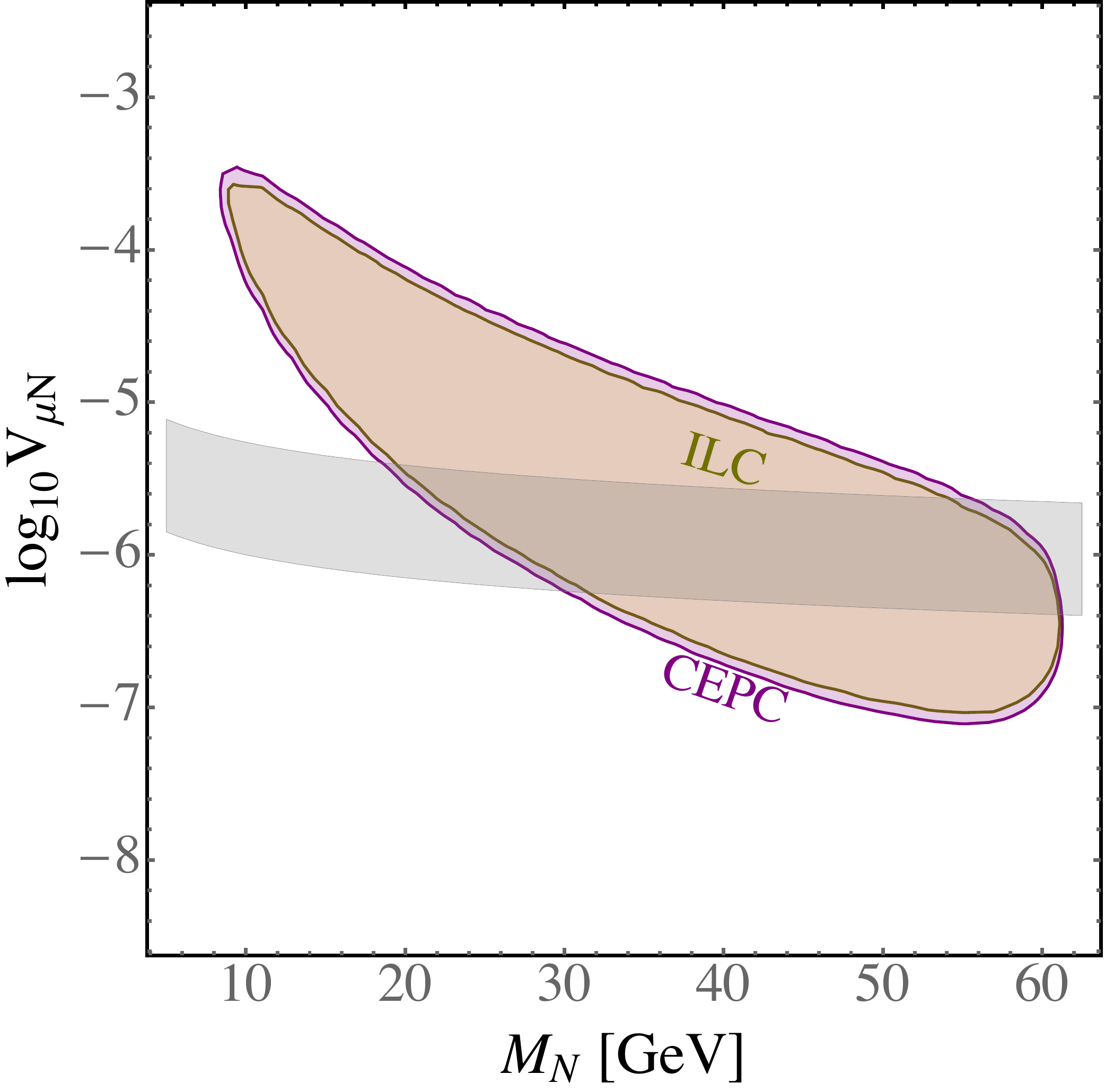}
\includegraphics[width=0.49\textwidth]{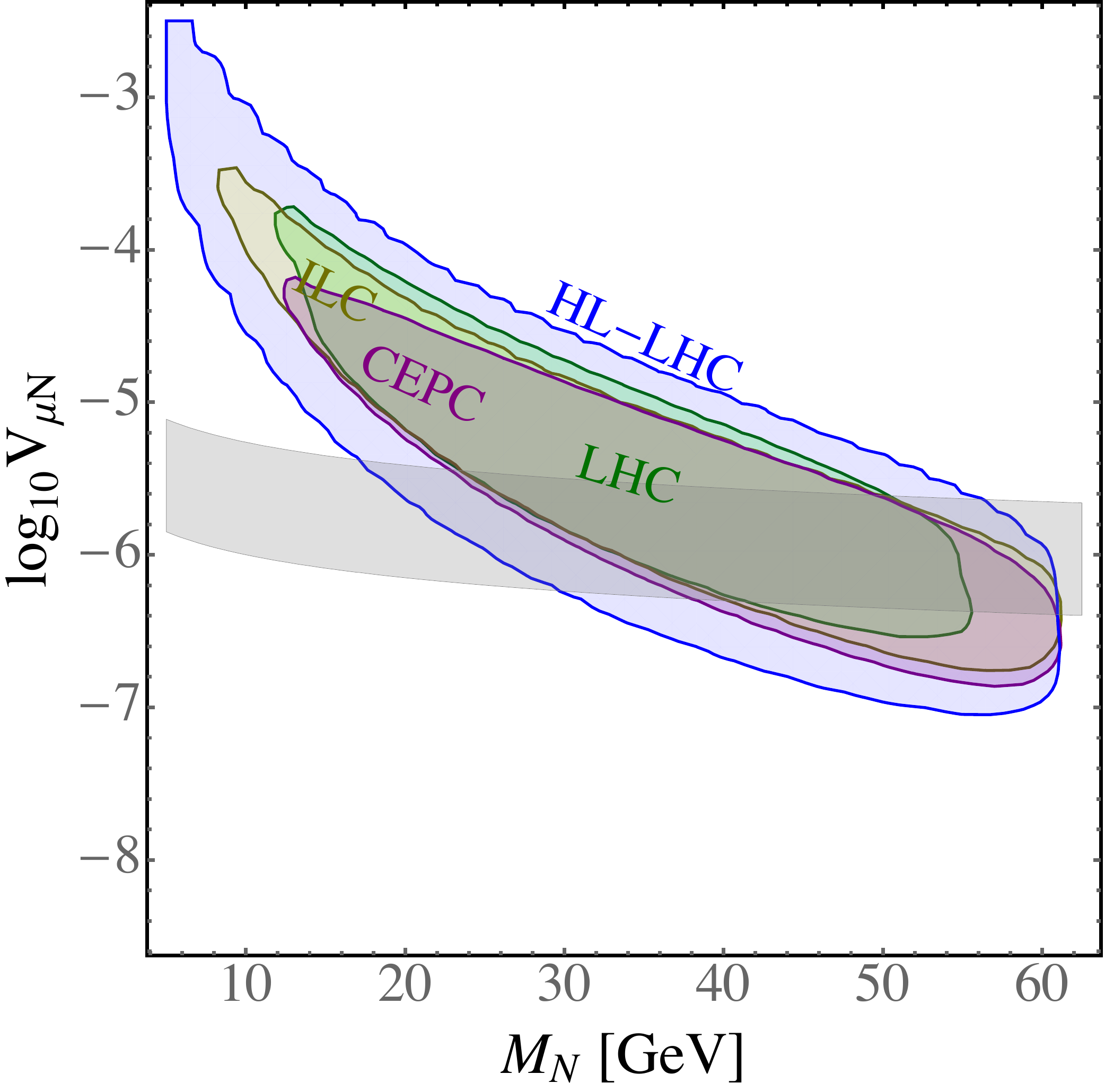}
\caption{Left: Excluded regions in the $(M_N, V_{\mu N})$ parameter space at 95\%~C.L. assuming no observation of a single displaced vertex for the 5000~fb$^{-1}$ ILC (red) and CEPC (blue).The grey band indicates the parameter region where a light neutrino mass in the interesting range is generated, $0.01~\text{eV} < m_\nu = V_{\mu N}^2 m_N < 0.3$~eV. Right: Excluded regions in the $(M_N, V_{\mu N})$ parameter space at 95\%~C.L. assuming no observation of two displaced vertices for the 100~fb$^{-1}$ LHC (green), 5000~fb$^{-1}$ ILC (red) and CEPC (purple) and 3000~fb$^{-1}$ HL-LHC (light blue).}
\label{fig:2DV}
\end{figure}
Generally, the leptonic colliders may be considered to have less background. However in this case, as we assume no background events at the LHC as well, this advantage is not realised. Because both the LHC and the proposed lepton collider detectors have similar dimensions and even as the lepton colliders have higher luminosities, this is cancelled by the lower cross section of the Higgs production processes. From Fig.~\ref{fig:LHC1DV} (left) and Fig.~\ref{fig:2DV} (left), it is evident that the 13~TeV LHC and the future colliders ILC/CEPC have a similar sensitivity reach down to $V_{\mu N} \sim 10^{-7}$ for RH neutrino mass $M_N \approx 55$~GeV. The ratio of the  relevant process cross sections almost equal the inverse of the luminosities. The high-luminosity run of LHC can further probe smaller mixing angles, as low as $10^{-8}$, for similar RH neutrino mass values. The sensitivity reach of MATHUSLA is similar to the HL-LHC due to its smaller geometric coverage.

In Fig.~\ref{fig:2DV}~(right), we show the sensitivity reach of the above collider options, demanding that the decays of the two heavy neutrinos generate \emph{two separate displaced vertices}. Since, this reduces the overall cross section and the probability, the effective event rate is severely reduced in this scenario, resulting in a more limited region of parameter space accessible at the 13 TeV LHC and other colliders. We stress that the two displaced vertices is a striking signature of the given model. Therefore, despite low event rate, this can serve as conclusive observational signal. We do not consider MATHUSLA for the two displaced vertex mode.

\begin{figure}[t!]
\centering
\includegraphics[width=0.49\textwidth]{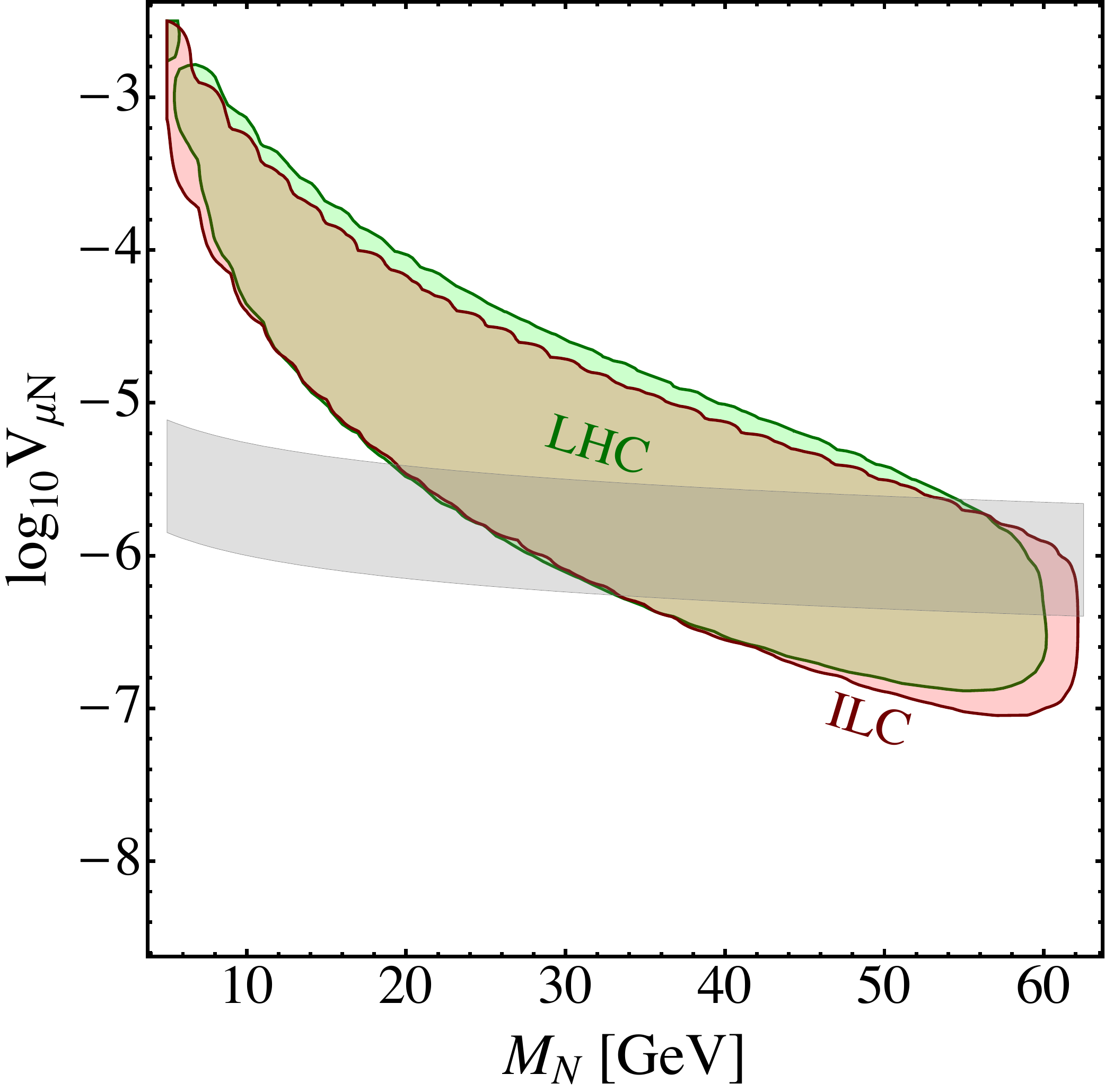}
\includegraphics[width=0.49\textwidth]{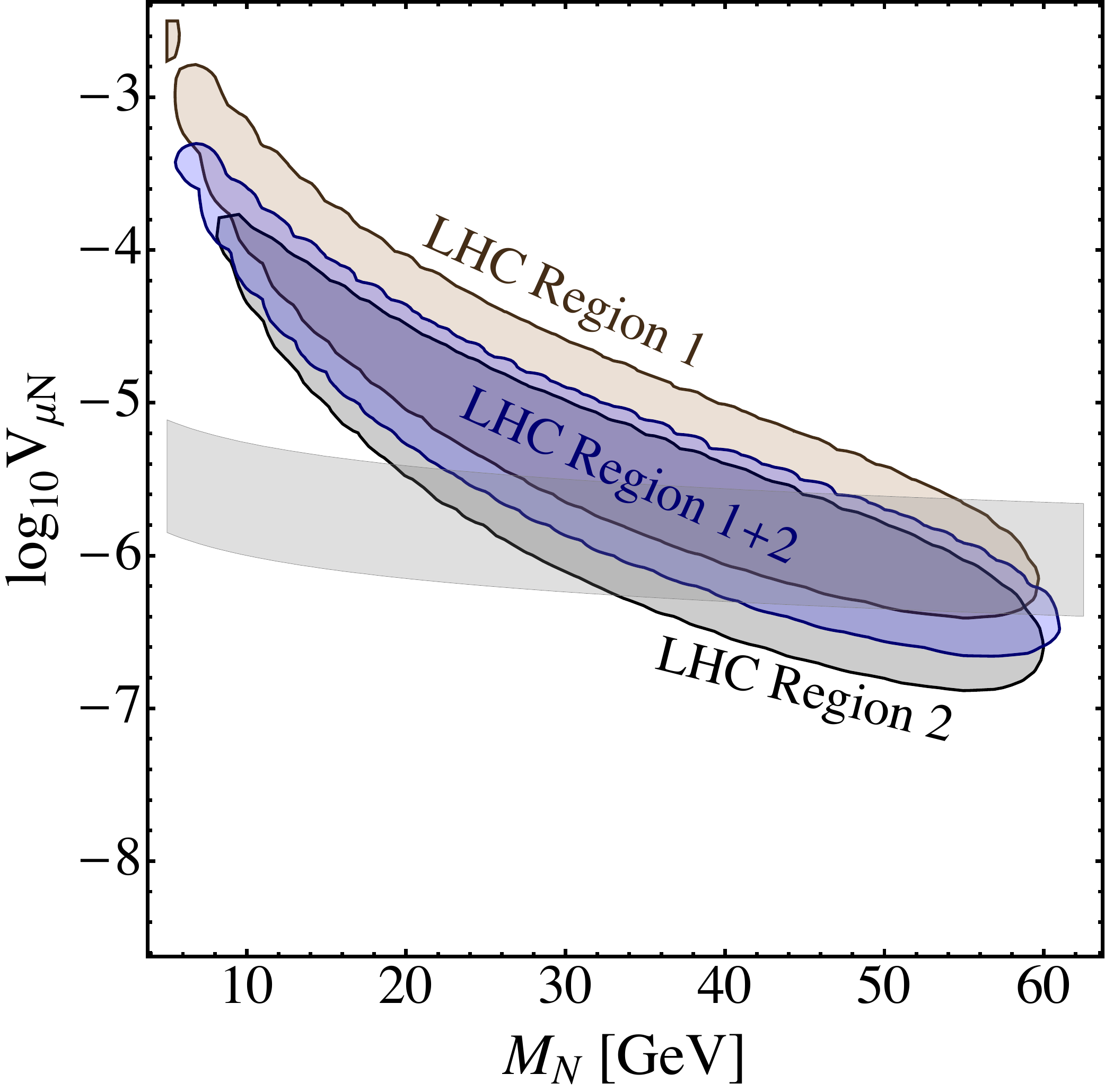}
\caption{Left: Viable parameter space assuming the observation of two events each containing two displaced vertices, at the LHC and the ILC. Right: Viable parameter space assuming {the observation of two events in the simplified LHC detector, each containing two displaced vertices where: (i) both events are fully in Region~1, (ii) both events are fully in Region~2 or (iii) one event is fully in Region~1 and the other fully in Region~2. In both plots, the Higgs mixing $\sin\alpha$ is not fixed but can vary up to its maximally allowed value.}}
\label{fig:region_compare}
\end{figure}
As an example how the actual observation of events would affect the model parameter space, we show in Fig.~\ref{fig:region_compare}~(left) the viable parameter space if two signal events each containing two displaced vertices are observed at the LHC and the ILC. {For a fixed Higgs mixing, the observation of a given number of events will generally correspond to a ring-like region. Instead, allowing $\sin\alpha$ to vary up to its experimental limit (i.e. not having measured it independently), but still fixing the number of displaced vertex events, one recovers a smaller region as shown in Fig.~\ref{fig:region_compare}~(left).} Such an observation could be supplemented by determining the RH neutrino mass through kinematical techniques to further constrain the parameter space. Lastly, the mass and especially the mixing strength directly affect the lifetime and thus the decay length of the RH neutrino. If a sufficient number of displaced vertices are observed, one can extract the lifetime from the exponential decay profile. An indication for this is provided by the region(s) of the detector we observe the signal events in. {Fig.~\ref{fig:region_compare}~(right) shows an example assuming that two signal events are observed in the simplified LHC detector, each containing two displaced vertices where: (i) both events are fully in Region~1, (ii) both events are fully in Region~2 or (iii) one event is fully in Region~1 and the other fully in Region~2. As before, we do not make an assumption on the value of the Higgs mixing, except that it is below its current experimental limit. As expected, observing events in the outer Region~2 probes smaller values of $V_{\mu N}$ compared to Region~1 and Fig.~\ref{fig:region_compare}~(right) illustrates how more detailed information on the displaced vertices can be used to constrain the parameter space in case of an observation. In addition to the three cases considered above, events in which the two displaced vertices are in different regions are possible as well.}

\section{Conclusion}
\label{cnclu}
In this work we have considered the $U(1)_{B-L}$ model and studied the SM Higgs decaying to two heavy RH neutrinos. The SM Higgs field mixes with the additional SM singlet Higgs that gives mass to the RH neutrinos through spontaneous symmetry breaking of the $B-L$ symmetry. In turn, the heavy neutrinos generate light neutrino masses through  seesaw mechanism. For RH neutrino masses $M_N < m_{h_1}/2$, the decay $h_1 \to N N$ is kinematically allowed and proportional to the mixing $\sin\alpha$ of the SM-like Higgs and gauge singlet Higgs. For such heavy neutrinos with masses $\lesssim 55$~GeV, their mixing with the light active neutrinos is expected to be of the order $V_{l N} \approx 10^{-6}$ for a standard Type-I seesaw scenario. This is far too low to produce heavy neutrinos in the leptonic charged current $p p \to W^{(*)} \to l N$. Besides, with the tight constraint on the mass of $Z^{\prime}$ gauge boson, $M_{Z^{\prime}} > 4.5$~TeV, from heavy resonance searches at the 13~TeV LHC, the pair-production cross-section of RH neutrinos through the $Z^\prime$ channel will be smaller. 

Instead, we investigate the production of RH neutrinos from Higgs decay $h_1 \to NN$  at the LHC and proposed future electron-positron colliders. The mixing between the SM-like Higgs and a heavy singlet Higgs is weakly constrained as $\sin\alpha < 0.3$. In the given model, this potentially allows for an abundant production of heavy neutrinos, that does not suffer any suppression due to very small active-sterile neutrino mixing. The heavy neutrino decay however depends crucially on the mixing parameter $V_{l N} \approx 10^{-6} - 10^{-2}$, leading to a potentially macroscopic decay length. The RH neutrinos in this scenario can be detected through displaced vertex searches at  colliders. In particular, we consider RH neutrino masses between 5~GeV to 62.5~GeV and simulate the rate of displaced neutrino events at the LHC, and the future colliders ILC and CEPC. Focussing on the coupling to the muon, we show that a sensitivity of $V_{\mu N} \approx 10^{-7}$ can be reached at the 13~TeV LHC with 100~$\text{fb}^{-1}$ both for a zero and non-zero background. For the later case, we estimate the background based on the LHC experimental searches on displaced vertex event rate.
 
For the lepton colliders ILC and CEPC, assuming zero background we arrive at a similar result. We note that the pair-production of RH neutrinos may also be used to constrain or determine the Higgs mixing angle $\sin\alpha$, although this of course requires the context of this model. For small active-sterile neutrino mixing, the RH neutrino production cross-section depends only on the Higgs mixing angle $\sin\alpha$ (and the RH neutrino mass $M_N$). In the limit of vanishing neutrino mixing, the RH neutrinos are invisible and escape as missing energy. Such as scenario is probed by invisible Higgs decay searches.

With the background of displaced vertex searches estimated to be either negligible (at the LHC and future electron colliders) or as projected from existing displaced vertex searches (at the LHC), we have shown that neutrino mixing strengths of order $V_{l N} \lesssim 10^{-6}$ can be probed for neutrino masses in the range $20~\text{GeV} \lesssim M_N \lesssim 60$~GeV. On the theoretical side, this impressive sensitivity, of the order expected to generate light Majorana neutrino masses $m_\nu \approx 0.1$~eV,  clearly hinges on the assumed large Higgs mixing of the production portal. It will need to be revised as future constraints become more severe. Experimentally, our simple background estimation requires verification and sophistication in a more detailed study. For example, we expect that the relevant background rate will strongly depend on the decay length and detector region of interest.

\section*{Acknowledgements}
MM acknowledges the support of the DST-INSPIRE research grant IFA14-PH-99, the Royal Society International Exchange program, and hospitality of  Institute for Particle Physics Phenomenology (IPPP), Durham University, UK and University College London, UK where the work has been initiated and part of the work has been carried out. WL acknowledges support via the China Scholarship Council grant CSC[2016]3100.

\end{document}